\documentclass[pre,superscriptaddress,longbibliography,footinbib,twocolumn,preprintnumbers,amsmath,amssymb]{revtex4-2}

\usepackage{amssymb}
\usepackage{amsmath}
\usepackage{comment} 

\usepackage{xr}
\makeatletter
\newcommand*{\addFileDependency}[1]{
  \typeout{(#1)}
  \@addtofilelist{#1}
  \IfFileExists{#1}{}{\typeout{No file #1.}}
}
\makeatother

\newcommand*{\myexternaldocument}[1]{%
    \externaldocument{#1}%
    \addFileDependency{#1.tex}%
    \addFileDependency{#1.aux}%
}

\myexternaldocument{supp}

\usepackage{xcolor}
\usepackage[pdftex]{graphicx}

\usepackage{empheq}

\usepackage{mathtools}

\usepackage{makecell}

\usepackage{siunitx}
\ifdefined\unit\else
  \ifdefined\NewCommandCopy
    \NewCommandCopy\unit\si
  \else
    \NewDocumentCommand\unit{O{}m}{\si[#1]{#2}}
  \fi
\fi

\usepackage{enumerate}
\usepackage{mathtools}
\usepackage{stmaryrd}

\usepackage{enumitem}

\usepackage{textcomp}
\usepackage{gensymb}

\newcommand{\ex}[1]{\mathrm{e}^{#1}}

\newcommand{\dd}[0]{\mathrm{d}}
\newcommand{\ii}[0]{\mathrm{i}}

\newcommand{\ee}[0]{\hat{\boldsymbol{e}}}
\newcommand{\rr}[0]{\boldsymbol{r}}

\newcommand{\PP}[0]{\boldsymbol{P}}

\newcommand{\qq}[0]{\boldsymbol{q}}

\newcommand{\kB}[0]{k_{\mathrm{B}}}

\newcommand{\uu}[0]{\hat{\boldsymbol{u}}}

\newcommand{\rotop}[0]{\boldsymbol{\mathcal{R}}}

\newcommand{\epsS}[0]{\varepsilon_S}

\definecolor{darkblue}{rgb}{0,0,0.6}
\definecolor{darkred}{rgb}{0.6,0,0}
\usepackage[colorlinks=true,urlcolor=darkblue,citecolor=darkblue,linkcolor=darkred]{hyperref}

\begin{document}

\title{Solvent-induced memory effects in a model electrolyte}

\author{Sleeba Varghese}
\affiliation{Sorbonne Universit\'e, CNRS, Physical Chemistry of Electrolytes and Interfacial Nanosystems (PHENIX), 4 place Jussieu, Paris, France}

\author{Benjamin Rotenberg}
\affiliation{Sorbonne Universit\'e, CNRS, Physical Chemistry of Electrolytes and Interfacial Nanosystems (PHENIX), 4 place Jussieu, Paris, France}
\affiliation{R\'eseau sur le Stockage Electrochimique de l'Energie (RS2E), FR CNRS 3459, 80039 Amiens Cedex, France}

\author{Pierre Illien}
\email{pierre.illien@sorbonne-universite.fr}
\affiliation{Sorbonne Universit\'e, CNRS, Physical Chemistry of Electrolytes and Interfacial Nanosystems (PHENIX), 4 place Jussieu, Paris, France}

\date{\today}

\begin{abstract}
The fluctuations of ions in polar solvents remain poorly understood theoretically due to the complex coupling between ionic motion and solvent polarization.
Indeed, while all-atom resolution can be achieved in numerical simulations, analytical approaches require suitable levels of coarse-graining. In this work, we describe ions and solvent molecules as interacting Brownian particles and use stochastic density functional theory to derive a generalized Langevin equation for the ionic charge density, explicitly accounting for solvent-mediated memory effects. 
In the regime where there is a clear timescale separation between fast solvent and slow ion dynamics, we obtain simple expressions for dynamical charge structure factors, which are validated by BD simulations. For slow solvents, we predict an emerging two-step relaxation in ionic dynamics. These results provide a mesoscopic approach for ion-solvent dynamics and open pathways to study fluctuation-induced phenomena in electrolytes.
\end{abstract}

\maketitle

\section{Introduction}

Electrolytes play a central role in a wide range of natural and technological processes, from biological systems to energy storage devices. While their static structure and transport properties has been successfully described thanks to classical theories developed for instance by Debye, Hückel and Onsager \cite{Harned1943,Resibois1968,Robinson2002}, capturing their dynamical properties—especially in the presence of explicit solvent fluctuations—remains a theoretical and computational challenge. Over the past decades, significant analytical progress has been made using dynamical density functional theories (DDFT) and mode-coupling theory (MCT) \cite{Yamaguchi2007,Roy2015, Bagchi1998, Chandra1999, Chandra2000a, Dufreche2002a,Dufreche2005a}, which have provided insights into ionic transport, diffusion, and collective dynamics.

More recently, stochastic density functional theory (SDFT), based on the Dean-Kawasaki equation \cite{Dean1996,Kawasaki1994,Illien2025a}, has emerged as a powerful framework to describe fluctuations in charged systems, offering both a microscopic foundation and a systematic way to derive effective theories. SDFT has successfully reproduced a broad range of dynamical properties in electrolytes, including ionic conductivity \cite{Demery2016,Avni2022,Avni2022a,Bonneau2023,Bonneau2024}, ionic self-diffusion \cite{Bernard2023a}, correlations \cite{Bonneau2025,Donev2019,Peraud2017}, noise in model nanopores \cite{Zorkot2016,Zorkot2016a,Zorkot2018}, and viscosity \cite{Robin2024}, while also enabling the study of fluctuation-induced phenomena such as nonequilibrium Casimir-like effects \cite{Dean2014a,Lu2015, Mahdisoltani2021a, Mahdisoltani2021c, Du2024,Du2025}.

However, a major limitation of existing theories is their treatment of the solvent as an implicit, structureless medium. This ignores its polar nature and the non-trivial coupling between solvent and ion dynamics. For instance, in aqueous electrolytes, the dielectric nature of water controls the structure and relaxation of ions \cite{Hubbard1977a, Buchner2009, Yang2023, Demery2026}, and, in turn, the presence of ions is known to significantly affect the structure and dynamics of water even at small concentrations \cite{Belloni2018, Jungwirth2018, Borgis2018, Duboisset2018, Duboisset2020}. While the dynamics of polar solvents alone have been addressed within dynamical theories \cite{Chandra1988,Bagchi1989,Vijayadamodar1989,Chandra1990a,Varghese2025,Dean2026}, the intertwined fluctuations of ions and solvent molecules, the importance of their respective timescales and of the related relaxation pathways, that have been the subject of many computational studies \cite{Schroder2008,Roy2010,Sega2013,Sega2014,Minh2023,Pireddu2024,Becker2025}, remain largely unexplored in analytical approaches.

{In this article}, we address these challenges by deriving a generalized Langevin equation obeyed by the ionic charge density, that accounts for solvent-induced memory effects. We deduce simple yet accurate expressions for dynamic charge structure factors in the limit of a fast solvent, and confront our theoretical predictions with Brownian dynamics (BD) simulations for a model electrolyte. Beyond this fast-solvent regime, we explore the emergence of novel dynamical behaviors when the solvent relaxation time becomes comparable to or even exceeds that of the ions, that may be relevant for electrolytes in organic solvents or under supercooled conditions.

{The article is organized as follows. In Sec.~\ref{sec_SDFT}, we introduce the model and the linearized SDFT equations obeyed by the ionic charge density and the solvent polarization; their derivation, and the approximations on which it relies, are detailed in Appendix~\ref{app_SDFT}. In Sec.~\ref{sec_GLE}, we derive a generalized Langevin equation for the ionic charge density. In Sec.~\ref{sec_fast}, we analyze the limit of fast-relaxing solvents and obtain simple expressions for the dynamic charge structure factors, which are compared to Brownian dynamics simulations in Secs.~\ref{sec_BD} and \ref{sec_DCSF}. Section~\ref{sec_slow} is devoted to slow-relaxing solvents, and we conclude in Sec.~\ref{sec_conclusion}. Analytical expressions, simulation details and additional numerical results are gathered in Appendices~\ref{sec_DCSF_general}--\ref{app_slow}.}

\section{Stochastic density functional theory}\label{sec_SDFT}

We consider a symmetric, binary electrolyte, in an unbounded three-dimensional volume. We denote by $\pm ze$ the charge of cations and anions (both with concentration $C_I$), and by $\rr_i^\pm(t)$ their positions at time $t$. These ions are embedded in explicit solvent `molecules', represented as point dipoles, of dipolar moment $p$ and concentration $C_S$. We denote by $\rr_j^S$ and $\uu_j$ the position and orientation of the $j$-th dipole. We assume that both ions and dipoles obey overdamped Langevin dynamics. {We denote by $D_I$ and $D_S$ the bare translational diffusion coefficients of ions and solvent molecules respectively, and by $D_S^r$ the rotational diffusion coefficient of solvent molecules}.

We introduce the microscopic densities of cations and anions $n_\pm(\rr,t) = \sum_i \delta(\rr-\rr_i^\pm(t))$ and that of solvent molecules $n_S(\rr,\uu,t) = \sum_j \delta(\rr-\rr_j^S(t)) \delta(\uu-\uu_j(t))$. The exact evolution equations for $n_\pm$ and $n_S$ were derived using It\^o's lemma in Ref. \cite{Illien2024e}, following the procedure that leads to Dean-Kawasaki equations or stochastic density functional theory \cite{Kawasaki1994,Dean1996,Illien2025a}.   {Details of the calculation are given in Appendix \ref{app_SDFT}}. We linearize these equations around a constant uniform state, i.e. $n_\pm = C_I + \delta n_\pm$ and $n_S = C_S/4\pi + \delta n_S$, and we find the following equations for the ionic charge density $\rho_I = ze(\delta n_+-\delta n_-)$:
\begin{equation}
    \partial_t \rho_I(\rr,t) = D_I \nabla^2 \rho_I + \varepsilon_0\kappa_I^2 {D_I} \nabla^2 \phi + \eta_I (\rr,t)
    \label{eq_LSDFT_I}
\end{equation}
{where $\phi(\rr,t)$ is the instantaneous electrostatic potential created by all the charges in the system (ions and solvent molecules), which is determined by Poisson's equation given below,} and {for the} solvent polarization field {$\PP(\rr,t) \equiv p\int \dd\uu \;  n_S(\rr,\uu,t)\uu= p\int \dd\uu \;  \delta n_S(\rr,\uu,t)\uu$:}
\begin{align}
    &\partial_t\PP(\rr,t)=D_S \nabla^2\PP -2D_S^r \PP \nonumber\\
    &\hspace{1cm }+\frac{p^2 C_S}{3 \kB T}{\nabla} [D_S \nabla^2\phi-2D_S^r \phi] + \boldsymbol{\eta}_S (\rr,t) \; .\label{eq_LSDFT_P}
\end{align}
{The range of validity of this linearization is discussed below [comment (iv)].} Here we introduced the inverse Debye screening length associated with the ions (in a medium of dielectric permittivity $\varepsilon_0$), $\kappa_I^2={2(ze)^2 C_I}/{\varepsilon_0\kB T}$ (solvent permittivity does not appear in this definition, as it comes into play explicitly through electrostatic interactions with solvent particles), and the Gaussian noises $\eta_I$ and $\boldsymbol{\eta}_S$ with zero average and correlations $\langle \eta_I(\rr,t)\eta_I(\rr',t') \rangle= -4(ze)^2 D_I C_I  \delta(t-t') \nabla^2 \delta(\rr-\rr')$ and $\langle \eta_{S,k}(\rr,t)\eta_{S,l}(\rr',t')\rangle = (2p^2 C_S /3)\delta_{kl}\delta(t-t') (-D_S\nabla^2+2D_S^r)\delta(\rr-\rr')$, respectively. These equations are completed with Poisson's equation, which reads $-\nabla^2\phi = (\rho_I+\rho_S)/\varepsilon_0$, where $\rho_S$ is the charge density associated with the solvent and $\varepsilon_0$ is the dielectric permittivity in vacuum. In the dipolar approximation, i.e. considering the system at distances much larger than the typical size of dipole, the solvent charge density simply reads $\rho_S(\rr,t) = -\nabla\cdot \PP(\rr,t)$. {This relation is not a definition of $\rho_S$: it follows from the multipole expansion of the charge distribution of each (neutral) solvent molecule, truncated at the dipolar order (see Appendix~\ref{app_SDFT}).} Taking the divergence of Eq.~\eqref{eq_LSDFT_P}, and in Fourier space (with the convention $\tilde\psi(\qq) = \int\dd\rr\; \ex{-\ii\qq\cdot\rr}\psi(\rr)$ for any space-dependent function), we find the following linear set of equations for $\tilde\rho_I$ and $\tilde\rho_S$ \cite{Illien2024e}:
\begin{equation}
    \partial_t \begin{pmatrix}
        \tilde \rho_I(\qq,t) \\
        \tilde \rho_S(\qq,t)
    \end{pmatrix}
=- \mathbf{M}
\begin{pmatrix}
        \tilde \rho_I(\qq,t) \\
        \tilde \rho_S(\qq,t)
    \end{pmatrix}
    +
    \begin{pmatrix}
        \Xi_I(\qq,t) \\
        \Xi_S (\qq,t)
    \end{pmatrix},
    \label{eq_main_result}
\end{equation}
with 
\begin{equation}
\label{ }
\mathbf{M} = 
\begin{pmatrix}
  D_I(q^2+\kappa_I^2)    &  D_I \kappa_I^2  \\
    D_S \kappa_S(q)^2   &    D_S(q^2+\kappa_S(q)^2) +2D_S^r
\end{pmatrix}.
\end{equation}
We introduced the wavenumber-dependent screening length associated with the charge polarization of the solvent $\kappa_S (q)^2  =\frac{2p^2 C_S }{3\varepsilon_0  \kB T a^2} \left(1+ \frac{q^2a^2}{2}\right) $ where $a\equiv(D_S/D_S^r)^{1/2}$ is the typical size of a solvent molecule, defined here through dynamical rather than structural properties. The noises $\Xi_I$ and $\Xi_S$ have zero average and correlations $\langle \Xi_\alpha (\qq,t) \Xi_\beta(\qq',t') \rangle = 2q^2 \varepsilon_0  \kB T D_\alpha \kappa_\alpha^2    (2\pi)^3 \delta (\qq+\qq') \delta(t-t') \delta_{\alpha\beta} $, for $\alpha\in\{I,S\}$.

Direct byproducts of Eq. \eqref{eq_main_result} are the charge-charge intermediate scattering functions of the model electrolyte, that are defined as $F_{\alpha\beta}(\qq,t) \equiv \langle \tilde\rho_\alpha(\qq,t)\tilde\rho_\beta(-\qq,0)\rangle/\sqrt{N_\alpha N_\beta}$, for $\alpha,\beta\in\{I,S\}$, and where $N_I$ and $N_S$ denote the total numbers of ions and solvent molecules, respectively. They typically read $F_{\alpha\beta}(\qq,t) = \sum_{\nu = \pm 1} \mathcal{F}_{\alpha\beta}^{(\nu)}(q)\ex{-\lambda_\nu |t|}$, where $\lambda_\nu$ are the eigenvalues of the matrix $\mathbf{M}$ and where the expressions of the weights $\mathcal{F}^{(\nu)}_{\alpha\beta}(q)$ were given in Ref.~\cite{Illien2024e}. We also define the charge-charge structure factors $S_{\alpha\beta}(\qq) =F_{\alpha\beta}(\qq,t=0)$ and dynamic structure factors $\tilde S_{\alpha\beta}(\qq,\omega) = \int_{-\infty}^\infty \dd t \; \ex{-\ii\omega t} F_{\alpha\beta}(\qq,t) $. Finally, it will be useful to consider the dynamic structure factors rescaled by their static counterpart, that we denote by $\mathcal{S}_{\alpha\beta}(\qq,\omega)\equiv \tilde S_{\alpha\beta}(\qq,\omega)/S_{\alpha\beta}(\qq) $.

The linear system of equations given in Eq. \eqref{eq_main_result} is the starting point of our analysis, and several comments follow:
(i)~We neglected short-range, steric interactions between the ions and solvent molecules, and we considered their evolution within the dipolar approximation. {Short-range repulsion is not included because it is not compatible with the linearization procedure, which assumes weak interactions, in the spirit of the Debye-H\"uckel approximation for electrostatic interactions [see comment~(iv)]: realistic steric repulsions must diverge at small distances, and therefore cannot be treated perturbatively. Soft-core repulsive interactions, which remain finite at short distances, could be included within the same framework, but we do not expect them to change the phenomenology discussed here, and this extension is beyond the scope of the present work.
These two approximations (no-short range repulsion and dipolar approximation) are only expected to be valid when considering the large-scale evolution of the system, i.e. for small enough wavevector  $|\qq|$. In this regime, the model captures the essential physics of ion-solvent interactions. For aqueous electrolytes, it agrees with simulations that represents explicitly solvent molecules \cite{Minh2023,Illien2024e}.}
(ii)~Without any ions, it is straightforward to show that the permittivity of the fluid of dipoles has the following Debye-like dependence on $\omega$: $\varepsilon(\omega)= \varepsilon_0+(\epsS-\varepsilon_0)/(1-\ii\omega/2D_S^r)$, where $\varepsilon_S = 1+\frac{p^2C_S}{3\varepsilon_0\kB T}$: this relates the effective static permittivity of the solvent in terms of the microscopic parameters of the model. We recently discussed the validity of such descriptions of polar solvents in the absence of ions, by systematic comparisons with BD simulations, and showed that the microscopic parameters describing the solvent could be rescaled using the Kirkwood factor (which is a measure of the local orientational correlations between neighboring dipoles) in order to account effectively for the short-range interactions between solvent particles \cite{Varghese2025}. This refinement is not necessary here but could be included in future work.  
{(iii)   The dynamic structure factors depend on both $D_S$ and $D_S^r$, i.e. on both translational and rotational properties of the solvent.}
{(iv)~The linearization of the SDFT equations around a uniform state is the dynamical counterpart of the Debye-H\"uckel (or random phase) approximation. It requires density fluctuations to be small compared to average densities. For the ions, this is ensured when electrostatic coupling is weak, namely when the Bjerrum length $\ell_B = (ze)^2/(4\pi\varepsilon_0\epsS\kB T)$ is small compared to the Debye length, or equivalently when $C_I\ell_B^3\ll1$. For the systems simulated in this work, $C_I\ell_B^3$ ranges from $3\times10^{-3}$ to $4\times10^{-2}$. Linearization does not capture strongly nonlinear effects such as ion pairing, which we account for effectively (Sec.~\ref{sec_DCSF}). For the solvent, the corresponding coupling parameter, $p^2C_S/(3\varepsilon_0\kB T)=\epsS-1$, is large for water-like solvents, and linearization amounts to a mean-field, Debye-like description of the dielectric response, which neglects local orientational correlations between dipoles; as shown in Ref.~\cite{Varghese2025}, this can be corrected by using effective parameters calibrated on simulations. Nonlinear couplings between density fluctuations, which would give rise to additional memory contributions of mode-coupling type~\cite{Bagchi1998,Chandra1999}, are therefore neglected in the present work, and the comparison with BD simulations (Sec.~\ref{sec_DCSF}) provides a direct test of this approximation.}

\section{Generalized Langevin description}\label{sec_GLE}

{Without further approximation on the dynamics of ions and solvent molecules, one can obtain from Eq.  \eqref{eq_main_result}  a closed equation for the ionic charge density.}
We solve for $\tilde \rho_S$ and reinject into the equation for $\tilde\rho_I$, which yields a generalized Langevin equation for the ionic charge density
\begin{equation}
			\partial_t \tilde\rho_I(\qq,t) = -\int_{-\infty}^t \dd t' \mathcal{M}(\qq,t-t') \tilde\rho_I(\qq,t') + \mathcal{X}(\qq,t)
            \label{eq_GLE_rhoI}
\end{equation}
with the memory kernel
\begin{eqnarray}
 \mathcal{M}(\qq,t) &=&  2D_I (q^2+\kappa_I^2)\delta(t) \nonumber\\
 && \hspace{0.5cm} - (D_I  \kappa_I^2)(D_S \kappa_S(q)^2)  \ex{-|t|/ \tau_S(q)}
\end{eqnarray}
and where the noise $\mathcal{X}$ is Gaussian, with zero average and correlations
\begin{align}
   & \langle \mathcal{X}(\qq,t)\mathcal{X}(\qq',t')  \rangle  = 2q^2\varepsilon_0 \kB T D_I \kappa_I^2 (2\pi)^3 \delta(\qq+\qq')  \nonumber\\
   &\times \left[ \delta(t-t')+\frac{\tau_S(q)}{2}D_I\kappa_I^2D_S\kappa_S(q)^2\ex{-|t-t'|/\tau_S(q)}\right].
\end{align}
We introduce the typical ($q$-dependent) relaxation time of the solvent: $   \tau_S(q) = [D_S(q^2+\kappa_S(q)^2)+2D_S^r]^{-1}$ \footnote{Note that $\tau_S$ is the \textit{longitudinal} relaxation time of the polarization field. Within our description and approximations, the \textit{transverse} relaxation time is simply given by $1/(D_S q^2+2D_S^r)$ \cite{Illien2024e,Varghese2025}}.
The ionic counterpart to this relaxation time is obtained as the inverse of the first diagonal term of $\mathbf{M}$, and reads $\tau_I(q)=[D_I(q^2+\kappa_I^2)]^{-1}$.
{Let us specify the status of Eq.~\eqref{eq_GLE_rhoI}. Given the linear system \eqref{eq_main_result}, it is exact: it is obtained by solving the equation for $\tilde\rho_S$, which is linear in $\tilde\rho_I$ and $\Xi_S$, and by substituting the solution into the equation for $\tilde\rho_I$. This is the projection of a linear system onto one of its variables, and no mode-coupling closure is involved: the exponential part of the memory kernel simply describes the relaxation of the solvent polarization, which is driven by the ions and acts back on them with a delay. In particular, no assumption is made on the relative magnitude of the relaxation times $\tau_I(q)$ and $\tau_S(q)$, so that Eq.~\eqref{eq_GLE_rhoI} can be used for both fast (Sec.~\ref{sec_fast}) and slow (Sec.~\ref{sec_slow}) solvents. Conversely, Eq.~\eqref{eq_GLE_rhoI} inherits all the approximations underlying Eq.~\eqref{eq_main_result}, namely overdamped dynamics, the dipolar approximation, the neglect of short-range interactions, and the linearization of the SDFT equations.}

\section{Timescale separation in fast-relaxing solvents}\label{sec_fast}

In aqueous electrolytes, a natural separation of timescales emerges because solvent and ionic motions are governed by fundamentally different physical mechanisms. Water relaxation is primarily a rotational process: individual molecules are small and can reorient through cooperative  rearrangements without requiring net mass transport. As a result, polarization fluctuations of the solvent occur on picosecond timescales \cite{Bakker2010,Laage2011}.  In contrast, ionic relaxation is controlled by translational diffusion, making them intrinsically much slower. Its characteristic time is typically estimated by $\varepsilon_S/(D_I\kappa_I^2)$, which is of the order of a few nanoseconds for a dilute ($\sim 0.1$ M) 1:1 electrolyte in water.

We then focus on the particular limit where the solvent can be assumed to relax much faster than ions. To this end, we define two typical relaxation times: $T_I ={\epsS}/{D_I\kappa_I^2}$ and $T_S={1}/{2 D_S^r \epsS}$,
which correspond respectively to the effective relaxation times of ions and to the effective rotational time of solvent molecules, that both involve the static permittivity $\epsS$. Importantly, in the limit $q\to 0$ and $T_S\ll T_I$, the two eigenvalues of the matrix $\mathbf{M}$ are given by $T_I^{-1}$ and $T_S^{-1}$, respectively. 

Considering the limit of $T_S\ll T_I$ in the generalized Langevin equation \eqref{eq_GLE_rhoI}  and using the general relation $\ex{-|u|/\epsilon} =_{\epsilon\to0} 2\epsilon \delta(u) + \mathcal{O}(\epsilon^2)$, we find that, at leading order in the small parameter $\zeta\equiv T_S/T_I$, the ionic density obeys $    \partial_t \tilde\rho_I(\qq,t) = -D_I [q^2+\kappa_\text{imp}^2] \tilde \rho_I(\qq,t)+ \tilde\eta_I(\qq,t)$, 
where {$\kappa_\text{imp}\equiv \kappa_I/\sqrt{\varepsilon_S}$} is the inverse screening length in a solvent of permittivity $\varepsilon_S$. This is the equation that would be obtained from the SDFT description of an electrolyte in an implicit solvent.

Going one step further in analyzing the regime $T_S/T_I \ll 1$, we now aim at finding corrections to the limit of an implicit solvent. Starting from the coupled equations for $\tilde\rho_I$ and $\tilde\rho_S$ [Eq. \eqref{eq_main_result}], we deduce  explicit yet lengthy expressions for the rescaled dynamical structure factors $\mathcal{S}_\alpha(\qq,\omega)$, that are written as the sum of two Lorentzian functions {(see Appendix~\ref{sec_DCSF_general})}. In the limit $T_S/T_I \ll 1$, we replace the eigenvalues $\lambda_{\pm 1}$ by $T_I^{-1}$ and $T_S^{-1}$ respectively, and expand the weight functions $\mathcal{F}^{(\nu)}_{\alpha\beta}(q)$ at leading order in $T_S/T_I$. In the $q\to 0$ limit, they take the simple form:
\begin{align}
    &\lim_{q\to 0}\mathcal{S}_{\alpha\beta}(q,\omega) \simeq 
    2 T_I \left[\frac{\mathcal{A}_{\alpha\beta}}{1+(\omega T_S)^2}
    +\frac{1}{1+(\omega T_I)^2} \right], \label{eq_S_q0}
\end{align}
where the coefficients $\mathcal{A}_{\alpha\beta}$ are the weight of solvent contributions and read:
$\mathcal{A}_{II}= \varepsilon_S(T_S/T_I)^3$;
$\mathcal{A}_{IS}= -(T_S/T_I)^2$ and 
$\mathcal{A}_{SS}= (T_S/T_I)/\varepsilon_S$.
In these three contributions, relaxation is dominated by the slow, ionic, mode of frequency $T_I$; but the solvent correction emerges at different order.
Equation \eqref{eq_S_q0} is important, as it provides simple estimates for the dynamical structure factors in terms of three parameters only: the two effective relaxation times $T_I$ and $T_S$, and the solvent relative permittivity $\epsS$. Similar but more complicated expressions are obtained for arbitrary $q$, and given explicitly in {Appendix~\ref{sec_DCSF_general}}.

\section{Brownian dynamics simulations}\label{sec_BD}

In order to determine the range of validity of our approximate analytical approach, we perform BD simulations of a more realistic  binary electrolyte using  LAMMPS \cite{Thompson2022}. 
The simulated system is identical to the one studied analytically, with the difference that we add steric (Lennard-Jones-like) interactions between all particles in the system. {These short-range repulsions are not included in our SDFT approach, but they are computationally necessary} to counterbalance attractive interactions of electrostatic origin and prevent particle overlap.

Simulations are performed for an electrolyte with ionic concentration $C_{I} = 0.1$ M and a water-like solvent at concentration $C_S = 55$ M. The solvent dipolar strength is $p = 1.85\ \mathrm{D}$, which corresponds to a Stockmayer solvent permittivity $\varepsilon_S = 147$ according to our previous study~\cite{Varghese2025}. To model transport properties, we use $D_{I} = 1.5 \times 10^{-9} \ \mathrm{m}^{2}\ \mathrm{s}^{-1}$ for the ionic translational diffusion coefficient, and $D_{S} = 2.3 \times 10^{-9} \ \mathrm{m}^{2}\ \mathrm{s}^{-1}$ and $D^r_{S} = 0.05\ \mathrm{ps}^{-1}$ ($a \simeq 2  $\si{\angstrom}) for the solvent translational and rotational diffusion coefficients, respectively. 
With these parameters, the typical relaxation times have orders of magnitude $T_I \sim 10^{-8}$ s and $T_S \sim 10^{-14}$ s, and are therefore well-separated. Further details on numerical simulations and additional results for other sets of parameters are provided in {Appendices~\ref{app_sim} and \ref{app_add}}.

\begin{figure}
\begin{tabular}{cc}
    \includegraphics[width=0.49\columnwidth]{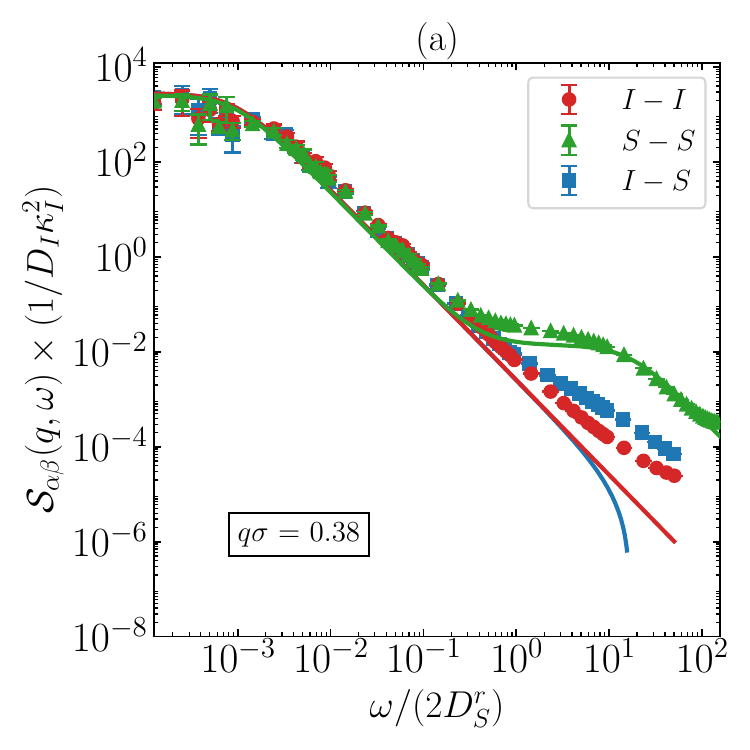}& 
     \includegraphics[width=0.49\columnwidth]{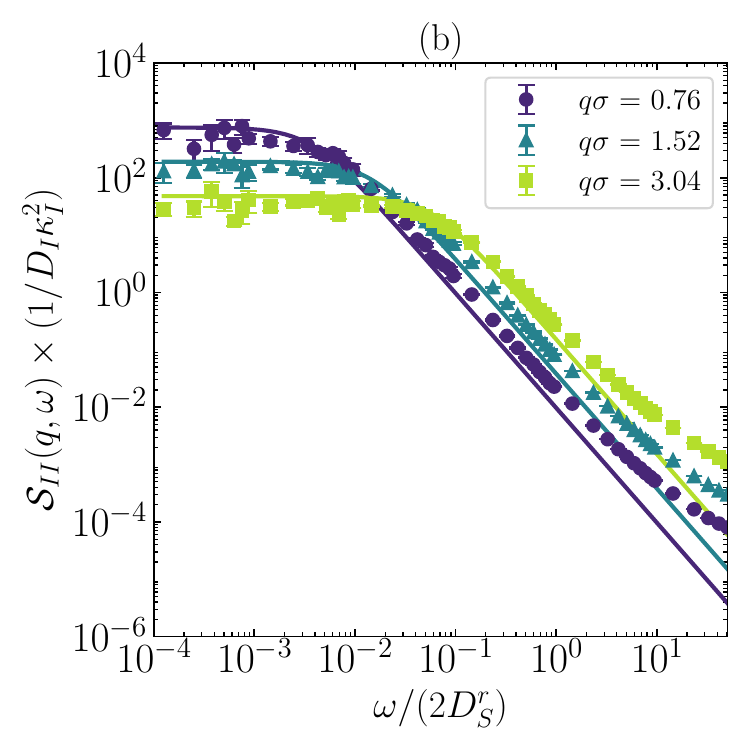}  \\
    \includegraphics[width=0.49\columnwidth]{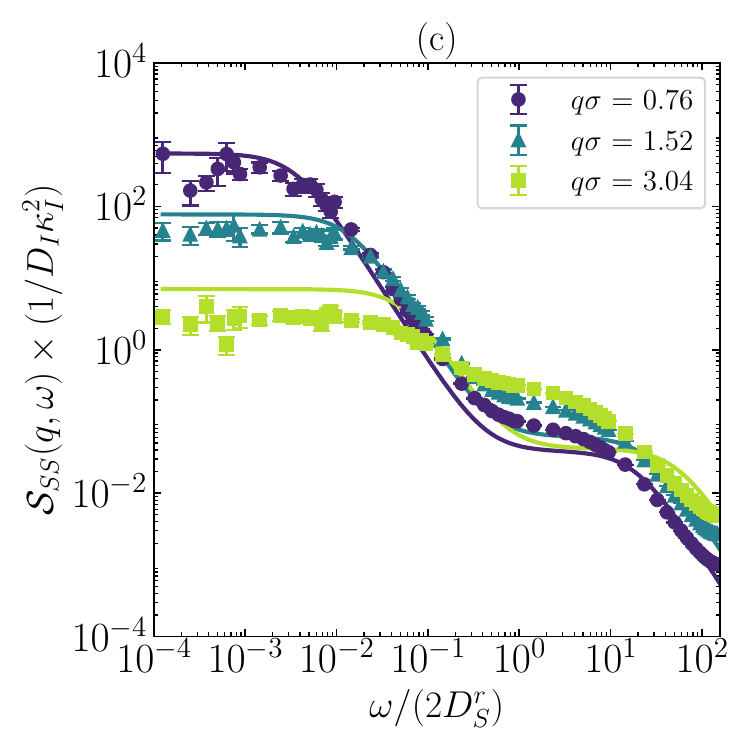} &  
     \includegraphics[width=0.49\columnwidth]{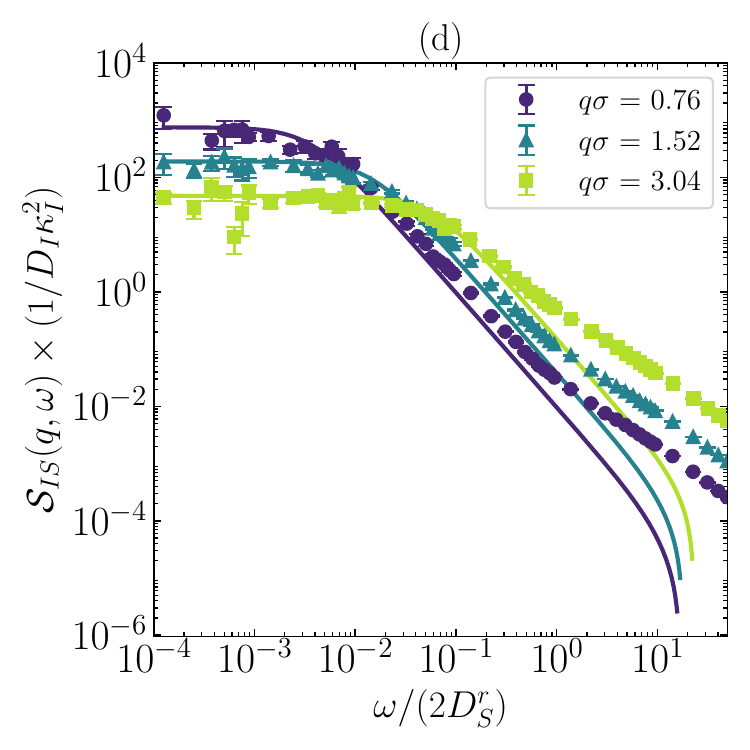}
\end{tabular}
    
    \caption{Dynamic charge structure factors $\mathcal{S}_{\alpha\beta}(\qq,\omega)$ measured in BD simulations (symbols) and calculated from SDFT (solid lines), (a) for the smallest wavevector accessible in the simulations
    and (b, c, d) for larger wavevectors (see {Appendix~\ref{sec_DCSF_general}} for analytical expressions). 
    Errorbars denote one standard error. 
    }
    \label{fig_DCSF_q0}
\end{figure}

\section{Dynamic charge structure factors}\label{sec_DCSF}

We first plot on Fig. \ref{fig_DCSF_q0}(a) the rescaled dynamic structure factors for the smallest wavevector accessible in numerical simulations, namely $q_\text{min}=0.38\sigma^{-1}$ ($\sigma = 3$\si{\angstrom} being the typical diameter of an ion or solvent particle in the simulations), and we use the expressions of $\mathcal{S}_{\alpha\beta}(q,\omega)$ for $q>0$ {(Appendix~\ref{sec_DCSF_general})} to obtain the effective values of $T_I$, $T_S$ and $\kappa_I$ (we choose here to set $\varepsilon_S$ to its previously determined value with these parameters~\cite{Varghese2025}, and details on the fitting procedure {are provided in Appendix~\ref{app_fit}}).

First, we observe that the BD data is well described by our model over many orders of magnitude. Second, we find an effective inverse Debye length $\kappa_I^\text{eff}=4.2$ nm$^{-1}$. This is to be compared to the bare value $(2C_I(ze)^2/\varepsilon_S\kB T)^{1/2} \simeq 9.5$ nm$^{-1}$. This means that the effective Debye length in the simulated electrolyte is apparently larger. This can be attributed to  ion pairing, which effectively reduces the concentration of free ions in simulations (see {Appendix~\ref{app_add}} for results on the cation-anion coordination number), and which is not accounted for in the theory. This effect is also reflected in the effective relaxation time of ions, estimated as $T_I^\text{eff}\simeq$ \num{1.5d-7}\unit{s}, which exceeds the bare value predicted theoretically. Finally, we find $T_S^\text{eff}\simeq$ \num{6d-13}\unit{s}, which is one order of magnitude higher than the bare theoretical value: this is attributed to local orientational correlations that are accounted for in the simulations, consistently with our previous results for a pure polar fluid \cite{Varghese2025}.

In order to probe the validity of our approach beyond the $q\to 0$ limit, we also confront our analytical estimates to numerical simulations for higher values of $q$ [Figs. \ref{fig_DCSF_q0}(b)-(d)]. We observe that SDFT predictions are more reliable in the low-frequency regime, while at high frequencies they only provide qualitative description of the underlying dynamics. {This discrepancy cannot be attributed to the overdamped approximation, since the BD simulations are themselves based on overdamped dynamics. Rather, at high frequencies, short-time dynamics is expected to be more sensitive to local steric and orientational correlations, which are present in the simulations but neglected in our linearized SDFT description. Nevertheless, Figs. \ref{fig_DCSF_q0}(b)-(d) show that for the low-frequency regime, our approach remains valid for much larger wavevectors, suggesting that slow relaxation modes are less sensitive to these microscopic details. This extends the applicability of our simplified description.}

\section{Towards slow-relaxing solvents}\label{sec_slow}

The results presented so far provide simple yet accurate estimates of the charge-charge correlation functions in conditions where the solvent relaxes much faster than ions. However, in ionic liquids \cite{Maity2025,Nakamoto2025} or under supercooled conditions \cite{Sha2019}, the timescales $T_S$ and $T_I$ are not expected to be well separated. We then explore the emergence of a `slow solvent' regime, and restrict ourselves to an analytical study, as molecular simulations become computationally expensive under these conditions.

\begin{figure}[t]
\centering
\includegraphics[width=0.49\columnwidth]{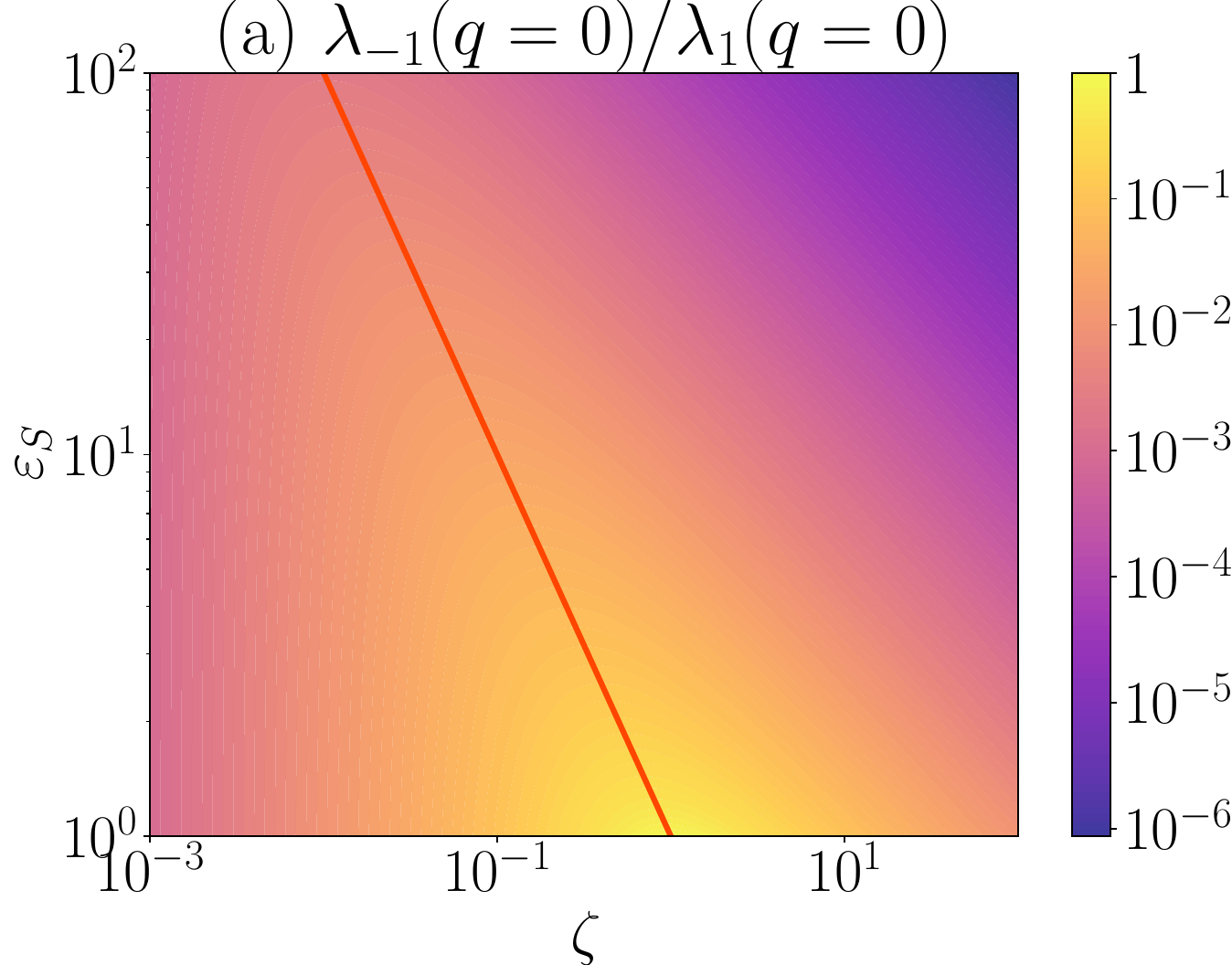}
\includegraphics[width=0.49\columnwidth]{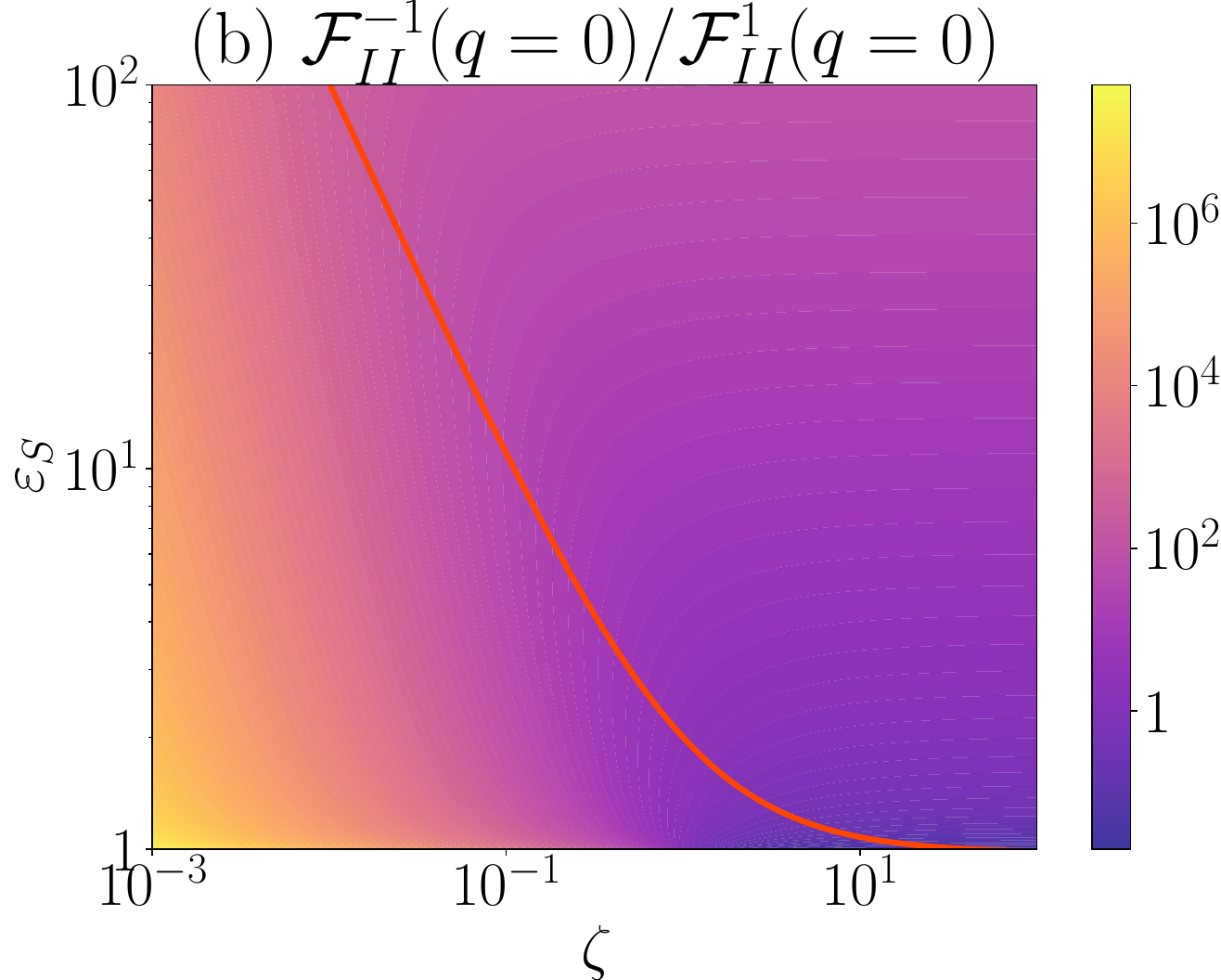}\\
\vspace{3pt}
\includegraphics[width=0.49\columnwidth]{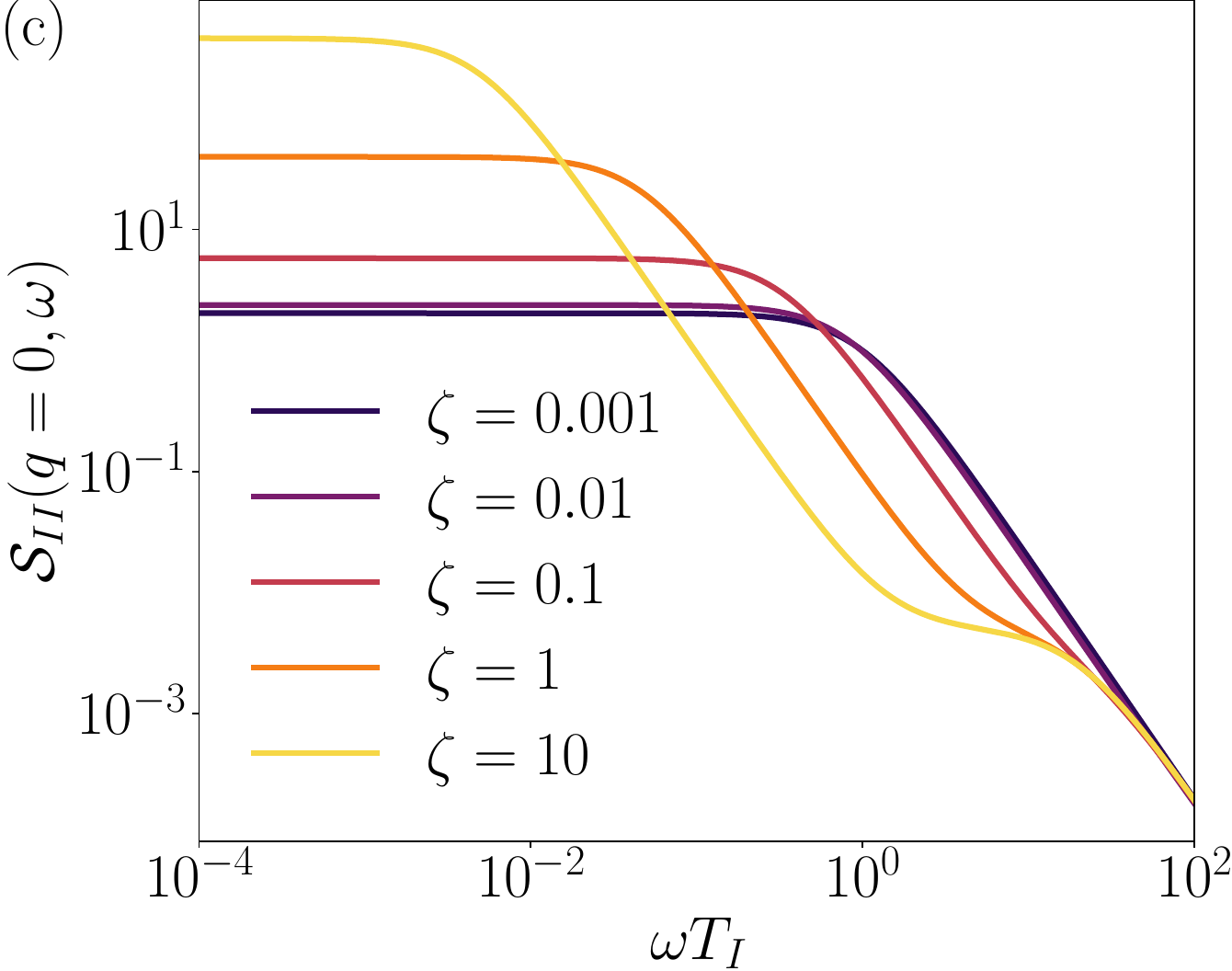}
\includegraphics[width=0.49\columnwidth]{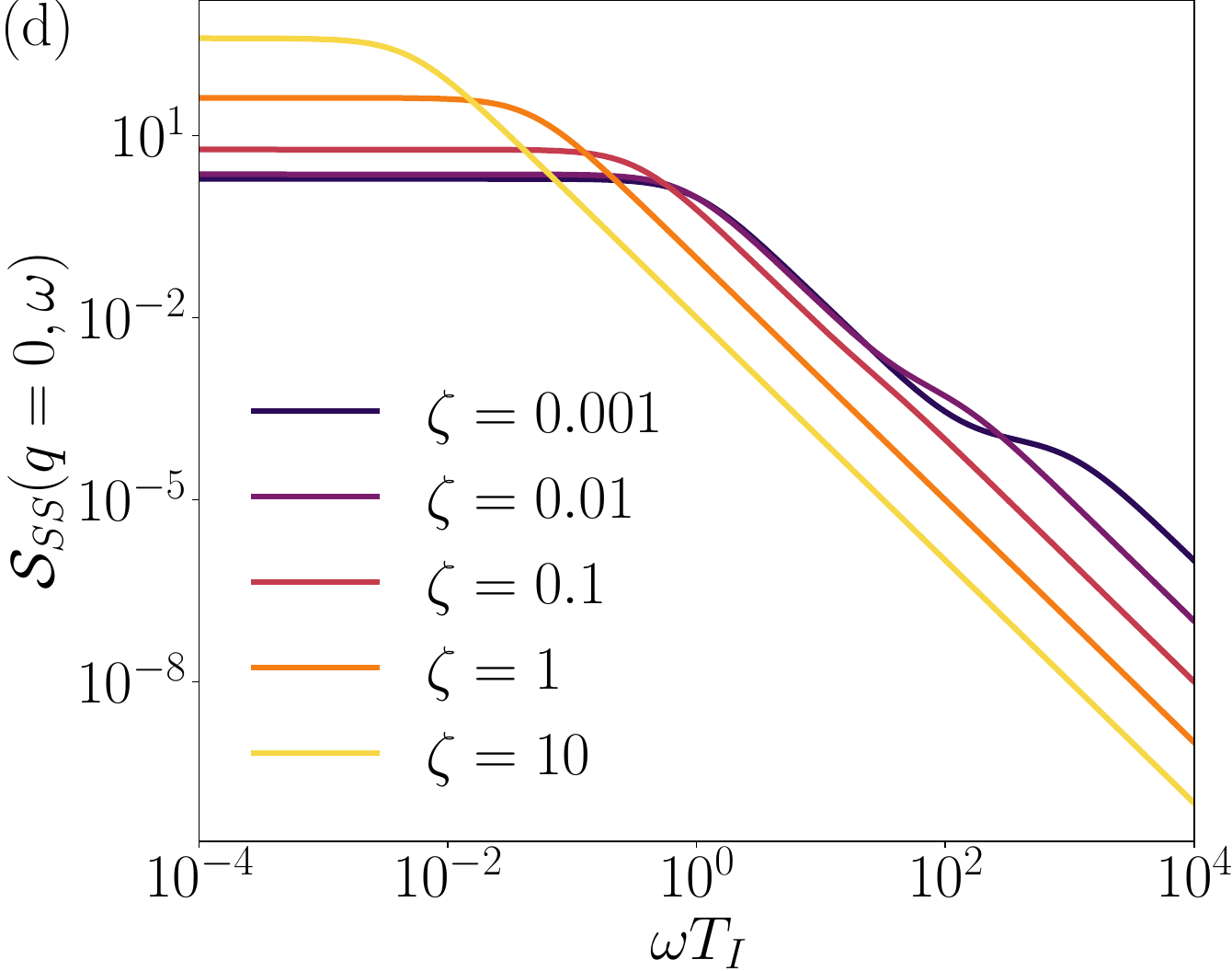}
\caption{
Ratio of eigenvalues $\lambda_{-1}(q=0)/\lambda_1(q=0)$ (a) and  weights $\mathcal{F}_{II}^{(-1)}(q=0)/\mathcal{F}_{II}^{(1)}(q=0)$ (b) of as a function of $\zeta = T_S/T_I$ and $\varepsilon_S$. The red line indicates $\varepsilon_S = 1/\zeta$ (a), and $\varepsilon_S = 1 + 1/\zeta$ (b).  Ionic (c) and solvent (d) rescaled dynamical structure factors for $\epsS=20$ and for varying $\zeta$. Ionic dynamics transitions from single-step relaxation ($\zeta \ll 1$) to two-step relaxation ($\zeta \gg 1$), and the opposite occurs for solvent dynamics.
}
\label{fig:relax_ion}
\end{figure}

To characterize the transition between the slow- and fast-solvent regimes, we first focus on the total relaxation time of the ions, defined as $\tau_{I}^\text{relax }(q) = [\int_{0}^{\infty}  F_{II}(q,t) \, \dd t]/S_{II}(q)$, and whose general expression is given in {Appendix~\ref{app_slow}}. In the $q\to0$ limit, it reads $\tau_{I}^\text{relax }(q=0)=T_I[1+(\epsS-1)\zeta]$, where $\zeta\equiv T_S/T_I$. As expected, in the limit of a fast-relaxing solvent ($\zeta \ll 1$), the ionic relaxation time is simply dominated by $T_I$. In the opposite limit of $\zeta\gg 1$, this relaxation time is related to the typical solvent timescale through the relation $\tau_{I}^\text{relax }(q=0)\simeq(\epsS-1) T_S $.

This opens the possibility for a two-step relaxation of the ion-ion dynamic charge structure factor. This can happen when the two contributions to the intermediate scattering function $\mathcal{F}_{II}^{(\pm 1)}$
occur at well-separated times but are of comparable magnitude. In order to determine the range of parameters where this may happen, we first compute the ratio between the two eigenvalues $\lambda_{-1}/\lambda_1$ {(Appendix~\ref{app_slow})}. For $q=0$, it reduces to ${(1 + \zeta \epsS - \sqrt{1 + \zeta^2 \epsS^2 + (2\epsS - 4)\zeta})^2}/(4\zeta)$. This expression is plotted as a function of $\zeta$ and $\epsS$ on Fig. \ref{fig:relax_ion}(a), and  is found to reach a maximum for $\epsS\zeta=1$. 
The two timescales are well separated provided that $\zeta \ll 1/\varepsilon_S$ or $\zeta \gg 1/\varepsilon_S$. Similarly, we compute the ratio between the two weights $\mathcal{F}_{II}^{(-1)}(q)/\mathcal{F}_{II}^{(1)}(q)$ {(Appendix~\ref{app_slow})} and plot it for $q=0$ as a function of $\zeta$ and $\epsS$ on Fig. \ref{fig:relax_ion}(b). We find that the two weights can be comparable if $\epsS$ is close enough to $1$ and $\zeta\gg1$. These observations indicate us the regime where two-step relaxation of ionic fluctuations may take place. We show an example of such an  unusually slow relaxation dynamics on Fig. \ref{fig:relax_ion}(c), and the corresponding solvent-solvent contribution on \ref{fig:relax_ion}(d).

\section{Conclusion}\label{sec_conclusion}

{In this article, we have studied the coupled dynamics of ions and solvent in a model electrolyte, where ions are point charges and solvent molecules are point dipoles, both undergoing overdamped translational and rotational Brownian motion. Starting from the linearized stochastic density functional theory for the ionic charge density and the solvent polarization, we derived a generalized Langevin equation for the ionic charge density, in which the coupling to the solvent results in a memory kernel and a colored noise. In the regime where the solvent relaxes much faster than the ions, we recovered the implicit-solvent description at leading order, and obtained simple expressions for the solvent contributions to the dynamic charge structure factors, which involve only the solvent permittivity and two relaxation times. The comparison with Brownian dynamics simulations showed that these expressions describe the simulated dynamics quantitatively at low frequencies, and qualitatively at high frequencies, provided that effective parameters are used to account for ion pairing and local orientational correlations between solvent molecules. Finally, in the opposite regime of slow solvents, we predicted a two-step relaxation of ionic charge fluctuations.
}

These findings open several avenues for future research. First, the validity of the timescale separation assumption should be systematically tested for a wider range of solvents, including those with slower relaxation dynamics. Second, the effective theory could be extended to account for spatial confinement, which is known to alter both the static and dynamic properties of electrolytes. Third, the predictions for the two-step relaxation and the slow mode could be confronted with experimental data, for example using dielectric spectroscopy or neutron scattering techniques. Such studies would not only deepen our understanding of the fundamental physics of electrolytes but also provide valuable insights for applications in energy storage and electrochemistry.

\section*{Acknowledgments}

The authors thank David Andelman for discussions. This project received funding from the European Research Council under the European Union’s Horizon 2020 research and innovation program (Grant Agreement No. 863473). P.I. acknowledges financial support from Agence Nationale de la Recherche through project TraNonEq (Grant No. ANR-24-CE30-0651).

{
\section*{Data availability}

The data that support the findings of this article are openly
available \cite{dataset}.

}

\appendix


\onecolumngrid

{

\section{Derivation of SDFT equations}
\label{app_SDFT}

In this Appendix, we detail the derivation of Eqs.~\eqref{eq_LSDFT_I} and \eqref{eq_LSDFT_P} of the main text, and we state explicitly the approximations on which it relies.

\subsection{Microscopic model}

We start from the overdamped Langevin equations obeyed by the positions of the ions $\rr_i^\pm$, the positions of the solvent molecules $\rr_j^S$ and their orientations $\uu_j$ ($|\uu_j|=1$):
\begin{align}
\frac{\dd \rr_i^\pm}{\dd t} &=\mp \mu_I ze\nabla\phi(\rr_i^\pm) + \sqrt{2D_I} \boldsymbol{\xi}_i^\pm(t), \label{langevin_rions}\\
\frac{\dd\rr^S_j}{\dd t} &=-\mu_S(p \uu_j\cdot \nabla) \nabla\phi(\rr_j^S) + \sqrt{2D_S} \boldsymbol{\xi}_j^S(t), \label{langevin_rs}\\
\frac{\dd \uu_j}{\dd t} &= \left\{-\mu_S^r p\,[\uu_j \times \nabla\phi(\rr_j^S)] + \sqrt{2D_S^r} \boldsymbol{\xi}_j^{S,r}(t) \right\}\times \uu_j, \label{langevin_ua}
\end{align}
where $\mu_I=D_I/\kB T$, $\mu_S=D_S/\kB T$ and $\mu_S^r=D_S^r/\kB T$ are the bare mobilities associated with the bare diffusion coefficients $D_I$, $D_S$ and $D_S^r$, respectively, and where the noises $\boldsymbol{\xi}_a^\alpha(t)$ are uncorrelated Gaussian white noises of zero average and unit variance, i.e. $\langle \xi_{i,n}^\alpha (t)\xi_{j,m}^\beta(t') \rangle = \delta_{\alpha\beta} \delta_{mn} \delta_{ij} \delta(t-t')$ ($n, m$ being components of the vectors).}{The first term in the right-hand side of Eq.~\eqref{langevin_rs} is the force exerted by the local electric field $-\nabla\phi$ on a point dipole $p\uu_j$, and the first term in the right-hand side of Eq.~\eqref{langevin_ua} corresponds to the rotation induced by the electric torque $p\uu_j\times(-\nabla\phi)$.\\

\paragraph{Electrostatic potential.} In Eqs.~\eqref{langevin_rions}-\eqref{langevin_ua}, $\phi(\rr,t)$ is the instantaneous electrostatic potential created by all the charges in the system, i.e. by the ions and by the solvent molecules. It is the solution of Poisson's equation
\begin{equation}
-\nabla^2\phi(\rr,t) = \frac{\rho_I(\rr,t)+\rho_S(\rr,t)}{\varepsilon_0},
\label{eq_Poisson_app}
\end{equation}
where $\rho_I(\rr,t)=ze[n_+(\rr,t)-n_-(\rr,t)]$ is the ionic charge density, with $n_\pm(\rr,t) = \sum_i \delta(\rr-\rr_i^\pm(t))$, and where $\rho_S$ is the charge density carried by the solvent molecules. Each solvent molecule is neutral and is described within the dipolar approximation, i.e. its charge distribution is replaced by the lowest nonvanishing term of its multipole expansion. The charge density associated with a point dipole $p\uu_j$ located at $\rr_j^S$ is $-p\,\uu_j\cdot\nabla\delta(\rr-\rr_j^S)$. Summing over all solvent molecules, we get
\begin{equation}
\rho_S(\rr,t) = -\nabla\cdot\PP(\rr,t), \qquad \PP(\rr,t)=p\sum_j \uu_j(t)\,\delta(\rr-\rr_j^S(t)) = p\int\dd\uu\; \uu\, n_S(\rr,\uu,t),
\label{eq_rhoS_app}
\end{equation}
where $n_S(\rr,\uu,t) = \sum_j \delta(\rr-\rr_j^S(t)) \delta(\uu-\uu_j(t))$ is the joint position-orientation density of solvent molecules. Therefore, the relation $\rho_S=-\nabla\cdot\PP$ is not a definition of $\rho_S$: it is a consequence of the dipolar approximation, and it is exact for point dipoles. Explicitly, the potential reads
\begin{equation}
\phi(\rr,t) = \frac{1}{4\pi\varepsilon_0}\int\dd\rr'\;\frac{\rho_I(\rr',t)-\nabla'\cdot\PP(\rr',t)}{|\rr-\rr'|},
\label{eq_phi_app}
\end{equation}
which is the continuous counterpart of Eq.~\eqref{eq_phi}, used in the simulations. \\

\paragraph{Interpretation of the rotational noise.} The noise in Eq.~\eqref{langevin_ua} is multiplicative, and its interpretation must be specified. We interpret it in the Stratonovich sense, which ensures that the norm $|\uu_j|=1$ is conserved and that, in the absence of electric torque, $\uu_j$ performs an isotropic Brownian motion on the unit sphere with rotational diffusion coefficient $D_S^r$. The derivation of the evolution equations for the densities relies on It\^o's lemma, and must therefore start from the It\^o form of Eq.~\eqref{langevin_ua}, which reads
\begin{equation}
\dd\uu_j = \left\{-\mu_S^r p\,[\uu_j \times \nabla\phi(\rr_j^S)]\times\uu_j - 2D_S^r\uu_j\right\}\dd t + \sqrt{2D_S^r}\,\dd\boldsymbol{W}_j\times\uu_j,
\label{langevin_ua_Ito}
\end{equation}
where $\boldsymbol{W}_j$ is a Wiener process, and where the noise-induced drift $-2D_S^r\uu_j$ (the factor $2=d-1$ being specific to three dimensions) is the It\^o correction. This choice has direct consequences on the dynamics, including at linear order: the noise-induced drift generates the rotational diffusion operator $D_S^r\rotop^2$ in Eq.~\eqref{DK3} below, and it is responsible for the relaxation term $-2D_S^r\PP$ in Eq.~\eqref{eq_LSDFT_P}. In the absence of electric torque, it yields $\langle\uu_j(t)\rangle = \ex{-2D_S^r t}\uu_j(0)$. A naive It\^o interpretation of Eq.~\eqref{langevin_ua}, without this drift, would conserve neither the norm of $\uu_j$ nor the isotropy of rotational diffusion. The noises in Eqs.~\eqref{langevin_rions} and \eqref{langevin_rs} are additive, and their interpretation is irrelevant.

}

{

\subsection{Dean-Kawasaki equations}

The equations satisfied by $n_\pm$ are obtained following the steps of Ref.~\cite{Dean1996} (see e.g. Refs.~\cite{Dean2014a,Demery2016} for early applications to charged systems). It is convenient to write them in terms of the charge density $\rho_I = ze (n_+ - n_-)$ and total number density $\mathcal{C} = n_+ + n_-$:
\begin{align}
    \partial_t \rho_I =& D_I\nabla^2 \rho_I+\mu_I (ze)^2 \nabla \cdot (\mathcal{C} \nabla \phi)+ze\nabla \cdot \left(\sqrt{2D_I n_+} \boldsymbol{\zeta}_{+}-\sqrt{2D_I n_-} \boldsymbol{\zeta}_{-}\right), \label{DK1}\\
    \partial_t \mathcal{C} =& D_I\nabla^2 \mathcal{C}+\mu_I  \nabla \cdot (\rho_I \nabla \phi)+\nabla \cdot \left(\sqrt{2D_I n_+} \boldsymbol{\zeta}_{+}+\sqrt{2D_I n_-} \boldsymbol{\zeta}_{-}\right),  \label{DK2}
\end{align}
where $\boldsymbol{\zeta}_{\pm}(\rr,t)$ are uncorrelated Gaussian white noises, of zero average and correlations $\langle\zeta_{\pm,n}(\rr,t)\zeta_{\pm,m}(\rr',t')\rangle=\delta_{nm}\delta(\rr-\rr')\delta(t-t')$.

The equation obeyed by the joint position-orientation density of solvent molecules $n_S(\rr,\uu,t)$ is obtained similarly, by applying It\^o's lemma to $\delta(\rr-\rr_j^S(t))\delta(\uu-\uu_j(t))$, with $\uu_j$ obeying Eq.~\eqref{langevin_ua_Ito}. Following Ref.~\cite{Illien2024e}, one gets
\begin{align}
        \partial_t n_S(\rr,\uu,t) =&  D_S\nabla^2_{\rr}n_S + D_S^r \rotop^2 n_S   +\nabla_{\rr} \cdot [n_S \mu_S p (\uu\cdot \nabla) \nabla\phi] +  \rotop\cdot [n_S\mu_S^r p\, \uu \times \nabla \phi] \nonumber\\
    &+\nabla_{\rr} \cdot\left(\sqrt{2D_S n_S } \boldsymbol{\zeta}_S\right)     +\rotop \cdot\left(\sqrt{2D_S^r n_S } \boldsymbol{\zeta}_{S,r}\right),  \label{DK3}
\end{align}
where $\rotop = \uu \times \partial_{\uu}$ is the rotational gradient operator~\cite{Doi1988,Dhont1996}, and where $\boldsymbol{\zeta}_S(\rr,\uu,t)$ and $\boldsymbol{\zeta}_{S,r}(\rr,\uu,t)$ are uncorrelated Gaussian white noises.

}

{

\subsection{Linearization}

We now linearize Eqs.~\eqref{DK1}-\eqref{DK3} around a constant and uniform state: $n_\pm = C_I + \delta n_\pm$ [so that $\rho_I=ze(\delta n_+-\delta n_-)$ and $\mathcal{C}=2C_I+c$] and $n_S = C_S/4\pi + \delta n_S$. In this state, $\phi=0$, so that $\phi$ is itself of first order in the fluctuations. The range of validity of this approximation is discussed in the main text [comment (iv) of Sec.~\ref{sec_SDFT}]. At linear order, the multiplicative noises become additive, so that their interpretation no longer matters, and we get:
\begin{eqnarray}
 \partial_t \rho_I(\rr,t) &=& D_I \nabla^2 \rho_I + \varepsilon_0\kappa_I^2 D_I \nabla^2 \phi + ze \sqrt{4D_I C_I}\, \eta_I (\rr,t), \label{lineq_rho}  \\
  \partial_t c(\rr,t) &=& D_I\nabla^2 c+ \sqrt{4D_I C_I}\, {\eta}_c(\rr,t), \label{lineq_c}  \\
 \partial_t \delta n_S(\rr,\uu,t) &=& D_S\nabla^2 \delta n_S + D_S^r \rotop^2 \delta n_S  + \mu_S p \frac{C_S}{4\pi} \nabla \cdot [(\uu\cdot \nabla)\nabla \phi]   + \mu_S^r p \frac{C_S}{4\pi} \rotop \cdot[\uu\times \nabla\phi] \nonumber\\
&&+\sqrt{2D_S} \sqrt{\frac{C_S}{4\pi}}\, \eta_S (\rr,\uu,t)+\sqrt{2D_S^r} \sqrt{\frac{C_S}{4\pi}}\, \eta_{S,r}(\rr,\uu,t),
    \label{cs_lin}
\end{eqnarray}
where we used $2\mu_I(ze)^2C_I=\varepsilon_0\kappa_I^2D_I$, by definition of the Debye screening length, and where we introduced the scalar noises $\eta_I=\nabla\cdot(\boldsymbol{\zeta}_+-\boldsymbol{\zeta}_-)/\sqrt{2}$, $\eta_c=\nabla\cdot(\boldsymbol{\zeta}_++\boldsymbol{\zeta}_-)/\sqrt{2}$, $\eta_S=\nabla\cdot\boldsymbol{\zeta}_S$ and $\eta_{S,r}=\rotop\cdot\boldsymbol{\zeta}_{S,r}$. They are uncorrelated, have zero average, and correlations
\begin{eqnarray}
    \langle \eta_{I}(\rr,t) \eta_{I}(\rr',t')\rangle = \langle \eta_{c}(\rr,t) \eta_{c}(\rr',t')\rangle
    &=&- \delta(t-t') \nabla_{\rr}^2\delta(\rr-\rr'),\\
    \langle \eta_S(\rr,\uu,t) \eta_S(\rr',\uu',t')\rangle
    &=&-\delta(t-t') \delta(\uu-\uu')\nabla_{\rr}^2\delta(\rr-\rr'),\\
    \langle \eta_{S,r}(\rr,\uu,t) \eta_{S,r}(\rr',\uu',t')\rangle 
    &=&-\delta(t-t') \delta(\rr-\rr') \rotop^2\delta(\uu-\uu').
\end{eqnarray}
Eq.~\eqref{lineq_c} is decoupled from the others and does not need to be considered. Eq.~\eqref{lineq_rho} is Eq.~\eqref{eq_LSDFT_I} of the main text. The evolution of the polarization $\PP=p\int\dd\uu\,\uu\,\delta n_S$ is obtained by multiplying Eq.~\eqref{cs_lin} by $p\uu$ and integrating over orientations. To this end, we use $\int\dd\uu\, u_k u_l = (4\pi/3)\delta_{kl}$, the fact that the Cartesian components of $\uu$ are eigenfunctions of $\rotop^2$ with eigenvalue $-2$ (they are $\ell=1$ spherical harmonics), and $\int\dd\uu\,u_k\rotop\cdot[\uu\times\nabla\phi] = -\int\dd\uu\,(\uu\times\ee_k)\cdot(\uu\times\nabla\phi) = -(8\pi/3)\partial_k\phi$, obtained by integration by parts on the unit sphere. This yields Eq.~\eqref{eq_LSDFT_P}, together with the expression of the noise correlations given in the main text. Finally, Eqs.~\eqref{eq_LSDFT_I} and \eqref{eq_LSDFT_P} are closed by Poisson's equation \eqref{eq_Poisson_app} with $\rho_S=-\nabla\cdot\PP$ [Eq.~\eqref{eq_rhoS_app}]. In Fourier space, $\tilde\phi(\qq,t) = [\tilde\rho_I(\qq,t)+\tilde\rho_S(\qq,t)]/(\varepsilon_0q^2)$, and taking the divergence of Eq.~\eqref{eq_LSDFT_P} leads to the closed linear system \eqref{eq_main_result} for $\tilde\rho_I$ and $\tilde\rho_S$.

}

\section{Analytical expressions of dynamical structure factors}

\label{sec_DCSF_general}

In this Appendix, we give the general expression of the rescaled dynamic charge structure factors $\mathcal{S}_{\alpha\beta}(\qq,\omega)$. For the sake of conciseness, we will express them in terms of the dimensionless wavevector $Q=qa$ . It is most convenient to write them as:
\begin{equation}
    \mathcal{S}_{\alpha\beta}(Q,\omega) = \sum_{\nu = \pm 1} \mathcal{W}_{\alpha\beta}^{(\nu)}(Q) \frac{1}{1+[\omega\lambda_\nu(Q)]^2} 
\end{equation}
where, in terms of the rescaled variables, the eigenvalues read:
\begin{equation}
T_I\lambda_{\pm1}(Q) = \frac{
 (Q^2 + 2\zeta \epsS + 2)K_I^2 + 2Q^2 \zeta \epsS \pm \Delta
}{
4K_I^2 \zeta
}
\end{equation}
where we introduced the dimensionless Debye length $K_I = \kappa_Ia$ and where we defined
\begin{equation}
\Delta =
\sqrt{
4\left(K_I^2+Q^2\right)^2 \epsS^2 \zeta^2
+K_I^4 (Q^2+2)(Q^2-8\zeta+2)
-4K_I^2\zeta (Q^2+2)(Q-K_I)(K_I+Q)\epsS
}.
\end{equation}
The weights $\mathcal{W}_{\alpha\beta}^{(\nu)}(Q)$ read:
\begin{eqnarray}
\mathcal{W}_{II}^{(1)}(Q)
&=&
\frac{2\zeta T_I}{\Delta (Q^2+2)}
\frac{\left(2\zeta\epsS(K_I^2+Q^2)\right)^2
-
\left(\Delta-K_I^2(Q^2+2)\right)^2}
{\Delta+K_I^2(Q^2+2)+2\zeta\epsS(K_I^2+Q^2)}
\\
\mathcal{W}_{II}^{(-1)}(Q)
&=&
\frac{2\zeta T_I}{\Delta (Q^2+2)}
\frac{\left(\Delta+K_I^2(Q^2+2)\right)^2
-
\left(2\zeta\epsS(K_I^2+Q^2)\right)^2}
{K_I^2(Q^2+2)+2\zeta\epsS(K_I^2+Q^2)-\Delta}
\\
\mathcal{W}_{SS}^{(1)}(Q)
&=&
\frac{T_I K_I^2}{\Delta (K_I^2+Q^2)\epsS}
\frac{K_I^4(Q^2+2)^2
-
\left(\Delta-2\zeta\epsS(K_I^2+Q^2)\right)^2}
{\Delta+K_I^2(Q^2+2)+2\zeta\epsS(K_I^2+Q^2)}
\\
\mathcal{W}_{SS}^{(-1)}(Q)
&=&
\frac{T_I K_I^2}{\Delta (K_I^2+Q^2)\epsS}
\frac{\left(\Delta+2\zeta\epsS(K_I^2+Q^2)\right)^2
-
K_I^4(Q^2+2)^2}
{K_I^2(Q^2+2)+2\zeta\epsS(K_I^2+Q^2)-\Delta}
\\
\mathcal{W}_{IS}^{(1)}(Q)
&=&
\frac{4\zeta T_I K_I^2
\left(\Delta-K_I^2(Q^2+2)-2\zeta\epsS(K_I^2+Q^2)\right)}
{\Delta\left(\Delta+K_I^2(Q^2+2)+2\zeta\epsS(K_I^2+Q^2)\right)}
\\
\mathcal{W}_{IS}^{(-1)}(Q)
&=&
\frac{4\zeta T_I K_I^2
\left(\Delta+K_I^2(Q^2+2)+2\zeta\epsS(K_I^2+Q^2)\right)}
{\Delta\left(K_I^2(Q^2+2)+2\zeta\epsS(K_I^2+Q^2)-\Delta\right)}
\end{eqnarray}


In the limit $T_S\ll T_I$, we get the approximated, $q$-dependent dynamic structure factors:
\begin{eqnarray}
    \mathcal{S}_{II}(q,\omega) &\simeq& 
    2T_I \left[\frac{8\epsS(\epsS-1)(q^2+\kappa_I^2/\epsS)}{\kappa_I^2(q^2a^2+2)^3}\left(\frac{T_S}{T_I}\right)^3\frac{1}{1+[\omega \bar\tau_S(q)]^2}+\frac{\kappa_I^2/\epsS}{ q^2+\kappa_I^2/\epsS}\frac{1}{1+[\omega \bar\tau_I(q)]^2} \right]\label{eq_SII_qnonzero}\\
    \mathcal{S}_{IS}(q,\omega) &\simeq& 
    2T_I \left[-\frac{4\epsS(q^2+\kappa_I^2/\epsS)}{\kappa_I^2(q^2a^2+2)^2}  \left(\frac{T_S}{T_I}\right)^2\frac{1}{1+[\omega \bar \tau_S(q)]^2}+\frac{\kappa_I^2/\epsS}{ q^2+\kappa_I^2/\epsS}\frac{1}{1+[\omega \bar\tau_I(q)]^2} \right]\label{eq_SIS_q_nonzero}\\
\href{}{}    \mathcal{S}_{SS}(q,\omega) &\simeq& 2T_I\left[ 
    \frac{2\epsS(q^2+\kappa_I^2/\epsS)}{(2+q^2a^2)(q^2+\kappa_I^2)\epsS}\left(\frac{ T_S}{T_I}\right)\frac{1}{1+[\omega \bar\tau_S(q)]^2}+\frac{\kappa_I^4(\epsS-1)}{\epsS^2(q^2+\kappa_I^2/\epsS)(q^2+\kappa_I^2)}\frac{1}{1+[\omega \bar\tau_I(q)]^2}\right] \label{eq_SS_qnonzero}
\end{eqnarray}
with 
\begin{equation}
    \bar\tau_S(q)= \frac{T_S}{1+q^2a^2/2} \qquad ; \qquad 
    \bar\tau_I(q)= \frac{T_I}{1+\epsS q^2/\kappa_I^2}.
\end{equation}
Finally, in the limit $q\to 0$, they reduce to the expression given in the main text.

\section{Numerical simulations and data analysis}\label{app_sim}

\subsection{Equations of motions}

In the overdamped limit, the dynamics of the ion positions $\rr_i^\pm$ ($N_I$ cations and $N_I$ anions), and positions and orientations of the $N_S$ solvent particles($\rr_j^S$ and $\uu_j$, respectively) are governed by
\begin{eqnarray}
    \frac{\dd\rr^{\pm}_i}{\dd t} &=& 
    -\mu_I  \sum_{\substack{k=1\\k\neq i}}^{N_I+N_S} \nabla_i U_{\alpha_i\alpha_k}^\text{rep}(|\rr_k-\rr_i|)\mp \mu_I (ze)\nabla\phi(\rr_i^\pm) + \sqrt{2D_{I}} \boldsymbol{\xi}^{\pm}_i(t), \label{eq_OLE_ri}\\
	\frac{\dd\rr^{S}_j}{\dd t} &=& -\mu_S  \sum_{\substack{k=1\\k\neq j}}^{N_I+N_S} \nabla_j U_{S\alpha_k}^\text{rep}(|\rr_k-\rr_j^S|)  -\mu_S (p\uu_j\cdot\nabla)\nabla\phi(\rr_j^S)
    +\sqrt{2D_{S}} \boldsymbol{\xi}^{S,t}_j(t), \label{eq_OLE_rs}\\
	 \frac{{\dd \uu}_j}{\dd t} &=& \left\{-[\mu_S^r p {\uu}_j \times \nabla\phi(\rr^{S}_j)] + \sqrt{2D_S^r} \boldsymbol{\xi}_j^{S,r}(t) \right\}\times {\uu}_j,\nonumber \\ \label{eq_OLE_us}
\end{eqnarray}
where $\alpha_i\in\{+,-,S\}$ denotes the species of particle $i$, and where $U_{\alpha\beta}^\text{rep}$ is  a Lennard-Jones potential:
\begin{align}
U_{\alpha\beta}^\text{rep}(r) =
\begin{cases}
4 \epsilon_{\alpha\beta}\Big[\Big(\frac{\sigma_{\alpha\beta}}{r}\Big)^{12}-\Big(\frac{\sigma_{\alpha\beta}}{r}\Big)^{6}\Big] 
& \textrm{if} \ r \leq r_\text{c}, \\
0 & \textrm{if} \ r > r_\text{c},
\end{cases} \label{eq_lj}
\end{align}
where $r_\text{c}$ is a cutoff radius.  The stochastic term $\boldsymbol{\xi}_a^\alpha(t)$ correspond to uncorrelated Gaussian white noises of zero average and unit variance, i.e. $\langle \xi_{i,n}^\alpha (t)\xi_{j,m}^\beta(t') \rangle = \delta_{\alpha\beta} \delta_{mn} \delta_{ij} \delta(t-t')$, where $n$ or $m$ are components of the vectors and Greek letters refer to particle labels. 
Finally, $\phi(\rr)$ is the electrostatic potential, which is obtained by solving Poisson's equation:
\begin{equation}
    -\nabla^2\phi(\rr) = \frac{\rho(\rr)}{\varepsilon_0} = \frac{\rho_I(\rr) + \rho_S(\rr)}{\varepsilon_0},
\end{equation}
where $\rho_I$ and $\rho_S$ are the charge density associated with ions and solvent, respectively. Within the dipolar approximation, it reads:
\begin{equation}
    \phi(\rr) = \frac{1}{4\pi \varepsilon_0}
    \left[
    \sum_{\epsilon=\pm}\sum_{i=1}^{N_I} \frac{\epsilon(ze)}{|\rr-\rr_i^\epsilon|}
    +\sum_{j=1}^{N_S} \frac{p\uu_j\cdot(\rr-\rr_j^S)}{|\rr-\rr_j^S|^3}
    \right].
    \label{eq_phi}
\end{equation}

\subsection{Simulation details}

Brownian dynamics simulations consist of a numerical integration of Eqs.~\eqref{eq_OLE_ri}-\eqref{eq_OLE_us}. To this end, we use the Large scale Atomic/Molecular Massively Parallel Simulator~\cite{Thompson2022}(LAMMPS) package, and the \texttt{brownian} package~\cite{Delong2015, Ilie2015}.
In addition to the system described in the main text (System I), we consider two additional sets of parameters (Systems II and III). All the parameters are summarized in Table~\ref{tab_systems}. 
In all cases, the temperature and solvent density are fixed at $T = 298\ \mathrm{K}$ and $C_{S} \simeq 0.033$ \unit{\angstrom}$^{-3}$ (55 M), respectively.

\begin{table*}
    \centering
    \begin{tabular}{c|c|c|c}
         \textbf{System label} &  \textbf{Molarity} & \textbf{Solvent dipole moment} & \textbf{Permittivity} \\
         \hline 
         I & 0.1 M (8) & 1.85 D & 146.5 \\
         II& 0.1 M (8) & 1.665 D & 64.07 \\ 
         III &0.4 M (30) & 1.85 D & 146.5 
    \end{tabular}
    \caption{Details of the representative systems used in this study. In the second column, values in parentheses denote the number of ion pairs. In the fourth column, the solvent permittivity values are taken from Ref~\cite{Varghese2025}.}
    \label{tab_systems}
\end{table*}

The Lennard-Jones and diffusion parameters for the ions are
\begin{eqnarray}
    \varepsilon_{II} &=& \epsilon_{++} = \epsilon_{--} = \epsilon_{+-} = 0.1\ \mathrm{kcal}\ \mathrm{mol}^{-1}\\
\sigma_{II} &=& \sigma_{++} = \sigma_{--} = \sigma_{+-} = 3.0\ \mathrm{\AA}, \\
D_{I} &=& D_{+} = D_{-} = 1.5 \times 10^{-9} \ \textrm{m}^{2}\ \textrm{s}^{-1}.
\end{eqnarray}
The Lennard-Jones interaction parameters for the Stockmayer fluid were taken from Ramirez {et al.}~\cite{ramirez2002density}, and the related simulation parameters are summarized in Table~\ref{tab_solv_para}. The ion-solvent cross-interaction parameters are determined using the Lorentz-Berthelot mixing rule.

\begin{table*}
    \centering
    \begin{tabular}{c|c|c}
         \textbf{Symbol} & \textbf{Definition} & \textbf{Value} \\
         \hline 
         $\sigma_{SS}$ & LJ diameter & $3.024\ \textrm{\AA}$ \\
         $\epsilon_{SS}$ & LJ energy & $0.4412\ \textrm{kcal}\ \textrm{mol}^{-1}$ \\ 
         $D_S$ & Solvent translational diffusion coefficient & $2.3 \times 10^{-9} \ \textrm{m}^{2}\ \textrm{s}^{-1}$ \\
         $D^{r}_S$ & Solvent rotational diffusion coefficient & $0.05\ \textrm{ps}^{-1}$
    \end{tabular}
    \caption{Simulation parameters for the Stockmayer solvent}
    \label{tab_solv_para}
\end{table*}

For all simulations, we use a cubic box of a side dimension $L\ =\ 50\ \mathrm{\AA}$, with PBC in all the coordinate directions. Thus, the smallest accessible wavevector, compatible with the PBC, is $2\pi/L = 0.126$\unit{\angstrom}$^{-1}$, which corresponds to $q_{\mathrm{min}}\sigma \simeq 0.38$.
Short-range steric interactions are computed using Eq.~\eqref{eq_lj}, with a real space cutoff $r_\text{c} = 7.56$ \unit{\angstrom}. Short-range electrostatic interactions are 
computed using the definition of the electrostatic potential [Eq. \eqref{eq_phi}], which combines contributions from ions and solvent molecules.
Long-range electrostatic interactions are computed using the particle-mesh Ewald (PME) algorithm~\cite{toukmaji2000efficient}, implemented in LAMMPS, with a relative root mean square error in the per-atom force calculations below $10^{-4}$.

All simulations employed a time step of 5 fs. 
For the systems with $C_I=0.1$ M, trajectories of $1\ \mu\mathrm{s}$ were generated, with observables sampled every 200 fs. For the system with $C_I=0.4$ M, simulations were run for a duration of 300 ns for four independent initial configurations, with data collected at the same sampling rate as the 0.1 M systems. All systems were initially equilibrated for a duration of 5 ns prior to data acquisition stage.

\subsection{Dynamic charge structure factors from simulations}
The partial Fourier components of the ionic charge density, and the longitudinal and transverse polarization densities, are sampled every 200 fs during the production run of the simulation, for selected wavevectors in the range $q\sigma \in [0.38,3.42]$, satisfying $q =  2\pi m / L$ with $m \in \mathbb{N}$ to ensure compatibility with periodic boundary conditions. The dynamic structure factor is computed for each wavevector using the Fast Fourier Transform algorithm, applied over $n$ blocks of the time series assumed to be statistically independent. For the system with $C_I=0.4$ M, the results are further averaged over the four independent initial conditions, with uncertainties estimated as the standard error across the independent trajectories.

\subsection{Fitting procedure}\label{app_fit}

The calibration proceeds as follows: (i) Perform regression fitting on the simulation data of longitudinal polarization density dynamic structure factor $\mathcal{S}_{SS}(q,\omega)$, for the smallest accessible wavevector ($q_\text{min}$). The fitting is carried out using the analytical expression of $\mathcal{S}_{SS}(q,\omega)$ in the $q\to0$ limit (see main text), which yields fit values for $T^{\mathrm{fit}}_S$, $T^{\mathrm{fit}}_I$ and $\varepsilon^{\mathrm{fit}}_S$. 
(ii) Next, we combine the $q=0$ and $q>0$ expressions for $\mathcal{S}_{SS}(q,\omega)$, to obtain the relation $1/\varepsilon^{\mathrm{fit}}_{S} = \frac{2(q^2+\kappa_I^2/\epsS)}{(2+q^2a^2)(q^2+\kappa_I^2)}$. This expression is then used to  determine $\kappa^{\text{eff}}_{I}$. In this step, as well as in the next, we set $\varepsilon_S$ to the permittivity of a pure solvent.
(iii) Finally, the optimal values for $T^{\text{eff}}_{S}$ and $T^{\text{eff}}_{I}$ are obtained from
$\bar\tau_S(q_{\text{min}}) = \frac{T^{\text{eff}}_S}{1+\left({q_{\text{min}}}a\right)^2/2}$ and $\bar\tau_I(q_{\text{min}}) = \frac{T^{\text{eff}}_I}{1+\epsS \left(q_{\text{min}}/\kappa_I^{\text{eff}}\right)^2}$, together with the conditions $\bar\tau_S(q_{\text{min}}) = T^{\mathrm{fit}}_{S}$ and $\bar\tau_I(q_{\text{min}}) = T^{\mathrm{fit}}_{I}$.

\section{Additional numerical results}\label{app_add}

\subsection{Dynamic charge structure factors: additional results}
\begin{figure}
\begin{tabular}{cc}
    \includegraphics[width=0.3\columnwidth]{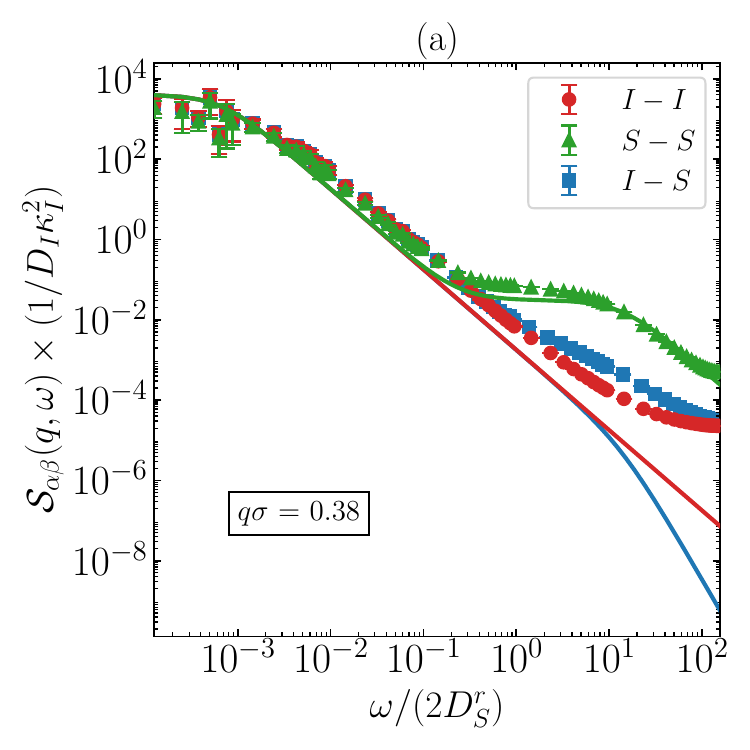} &
    \includegraphics[width=0.3\columnwidth]{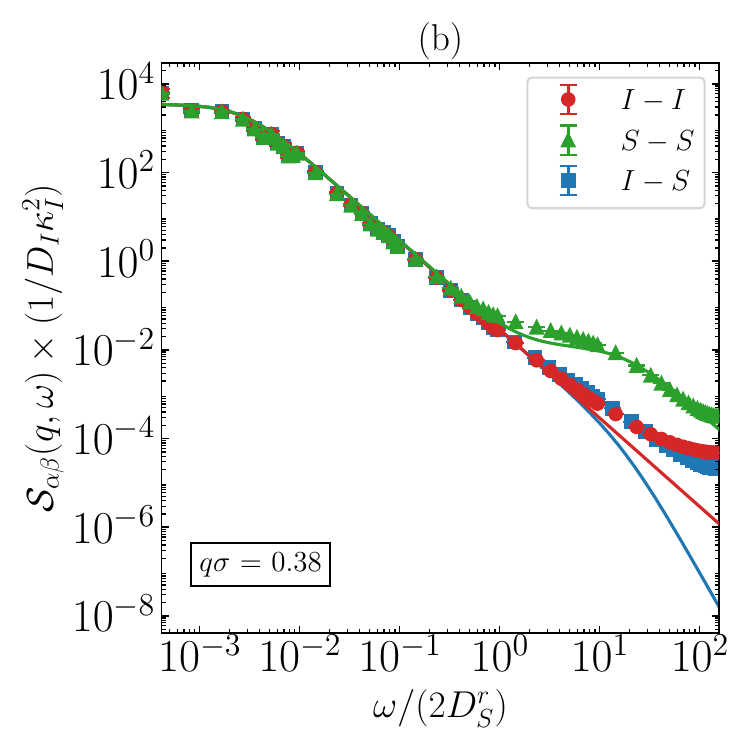} 
\end{tabular}
    \caption{Dynamic charge structure factors $\mathcal{S}_{\alpha\beta}(\qq,\omega)$ measured in Brownian dynamics simulations (symbols) and calculated from SDFT (solid lines), for the smallest wavevector accessible in the simulations. The SDFT estimates are obtained using the $q=0$ expressions [see main text].
    The values for the fit parameters $(\varepsilon^{\mathrm{fit}}_S, T^{\mathrm{fit}}_{S}, T^{\mathrm{fit}}_{I})$, for Systems II (a)  and III (b) are $(6.26,\ 0.714\ \mathrm{ps},\ 1.50\times 10^{4}\ \mathrm{ps})$ and $(76.6,\ 0.546\ \mathrm{ps},\ 3.472\times 10^{3}\ \mathrm{ps})$, respectively. The errorbars denote one standard error.
    }
    \label{fig_DCSF_q0_SM}
\end{figure}

\begin{figure*}
\begin{tabular}{ccc}
    \includegraphics[width=0.33\linewidth]{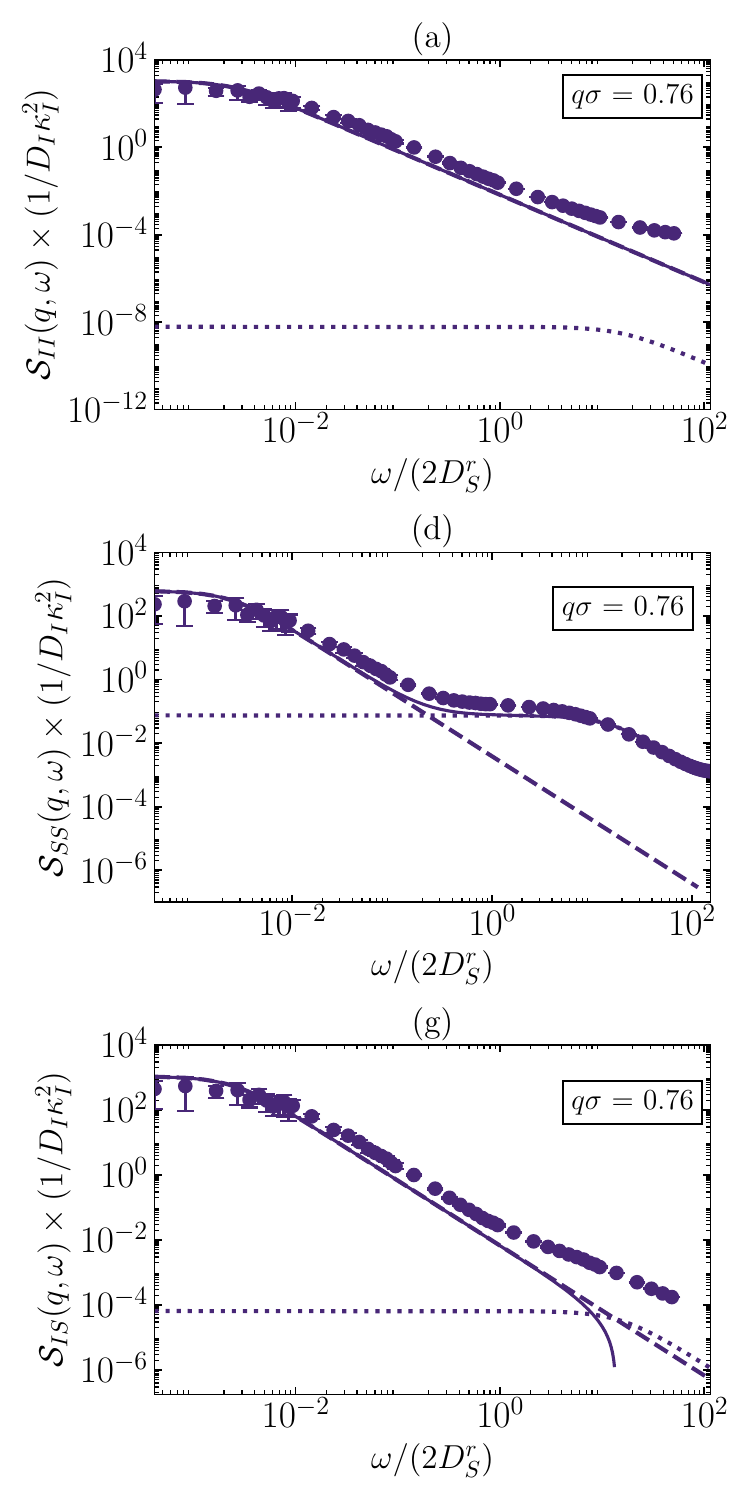} &
    \includegraphics[width=0.33\linewidth]{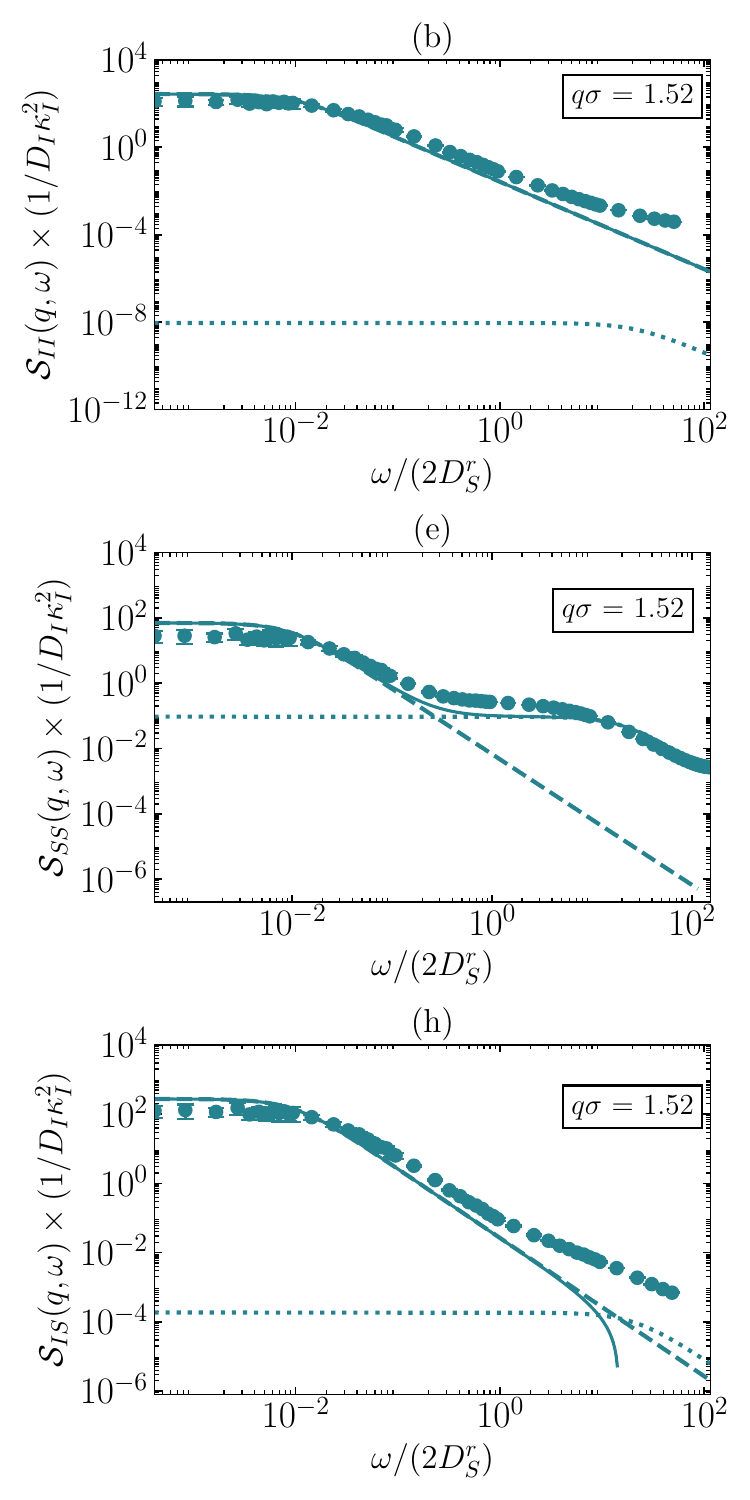} &
    \includegraphics[width=0.33\linewidth]{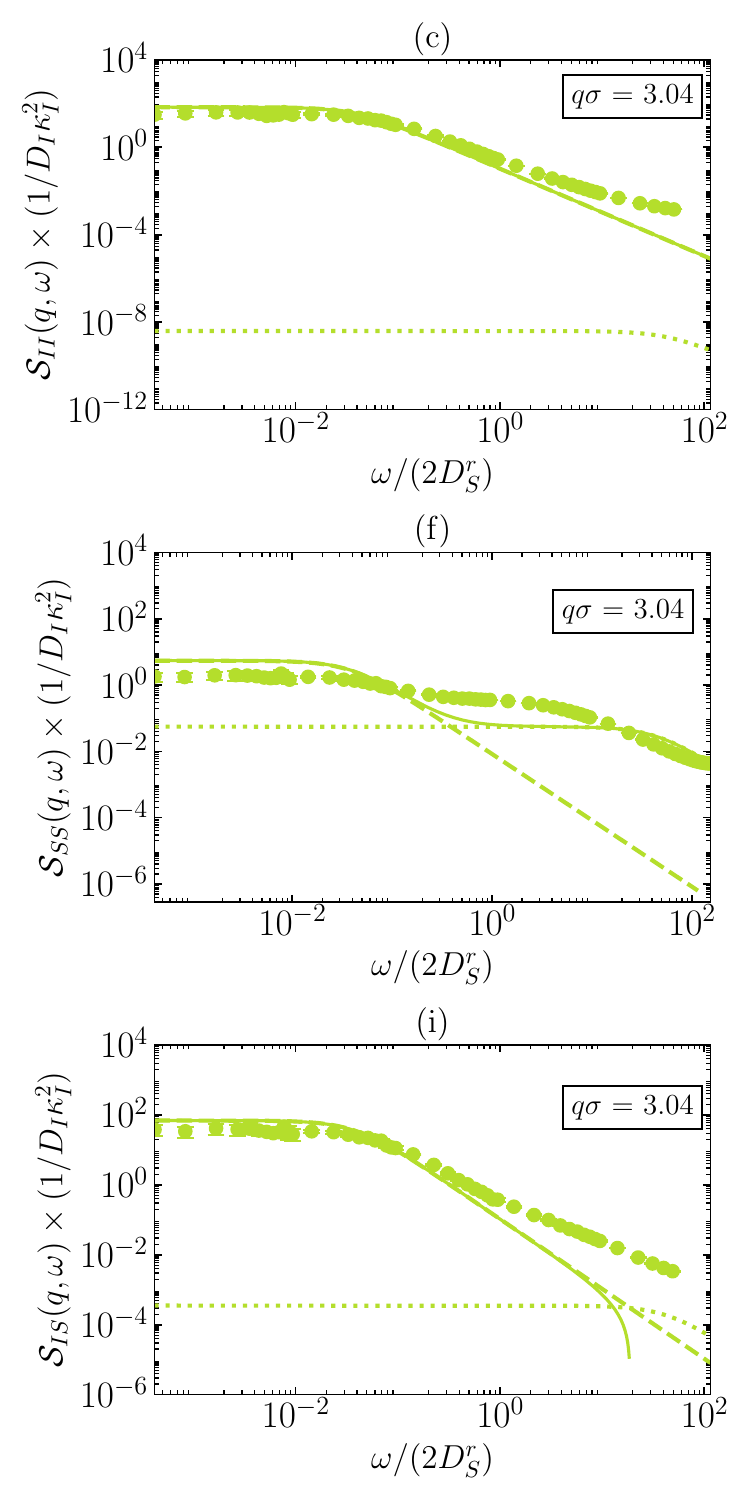} 
\end{tabular}
 \caption{Dynamic charge structure factors $\mathcal{S}_{\alpha\beta}(\qq,\omega)$ at larger wavevectors for   System II, calculated from Brownian dynamics simulations (symbols) and SDFT (lines). Solid lines show the full SDFT predictions based on Eqs.~\eqref{eq_SII_qnonzero}-\eqref{eq_SS_qnonzero}, while dashed lines denote contribution from their second term only. In plots (a)-(f) dotted lines represent the contribution from the first term of Eqs.~\eqref{eq_SII_qnonzero} and \eqref{eq_SS_qnonzero}, whereas in plots (g)-(i) they correspond to the opposite of the first term of Eq.~\eqref{eq_SIS_q_nonzero}. 
 The errorbars denote one standard error.}
 \label{fig_DCSF_q_nonzero_SM_p0_9}   
 \end{figure*}

\begin{figure*}
\begin{tabular}{ccc}
    \includegraphics[width=0.33\linewidth]{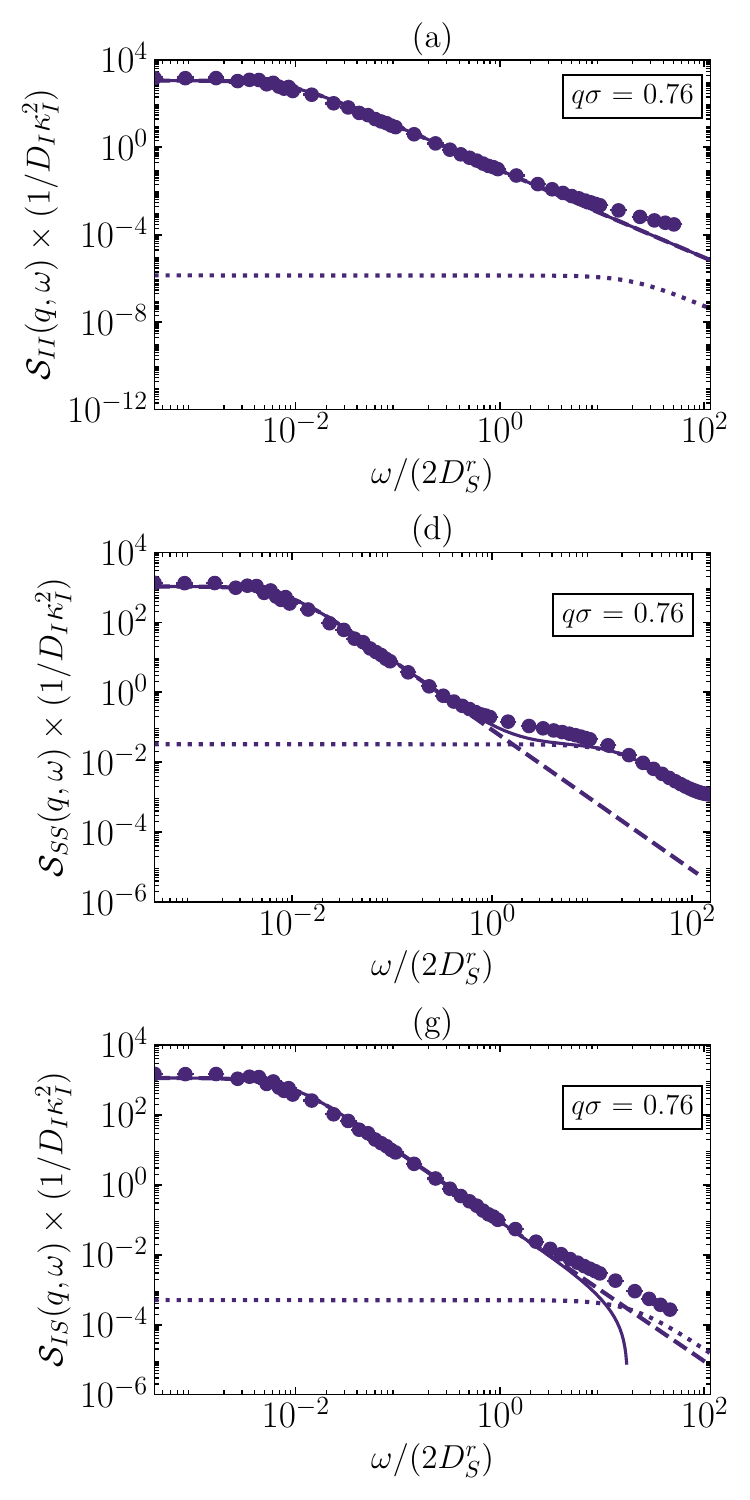} &
    \includegraphics[width=0.33\linewidth]{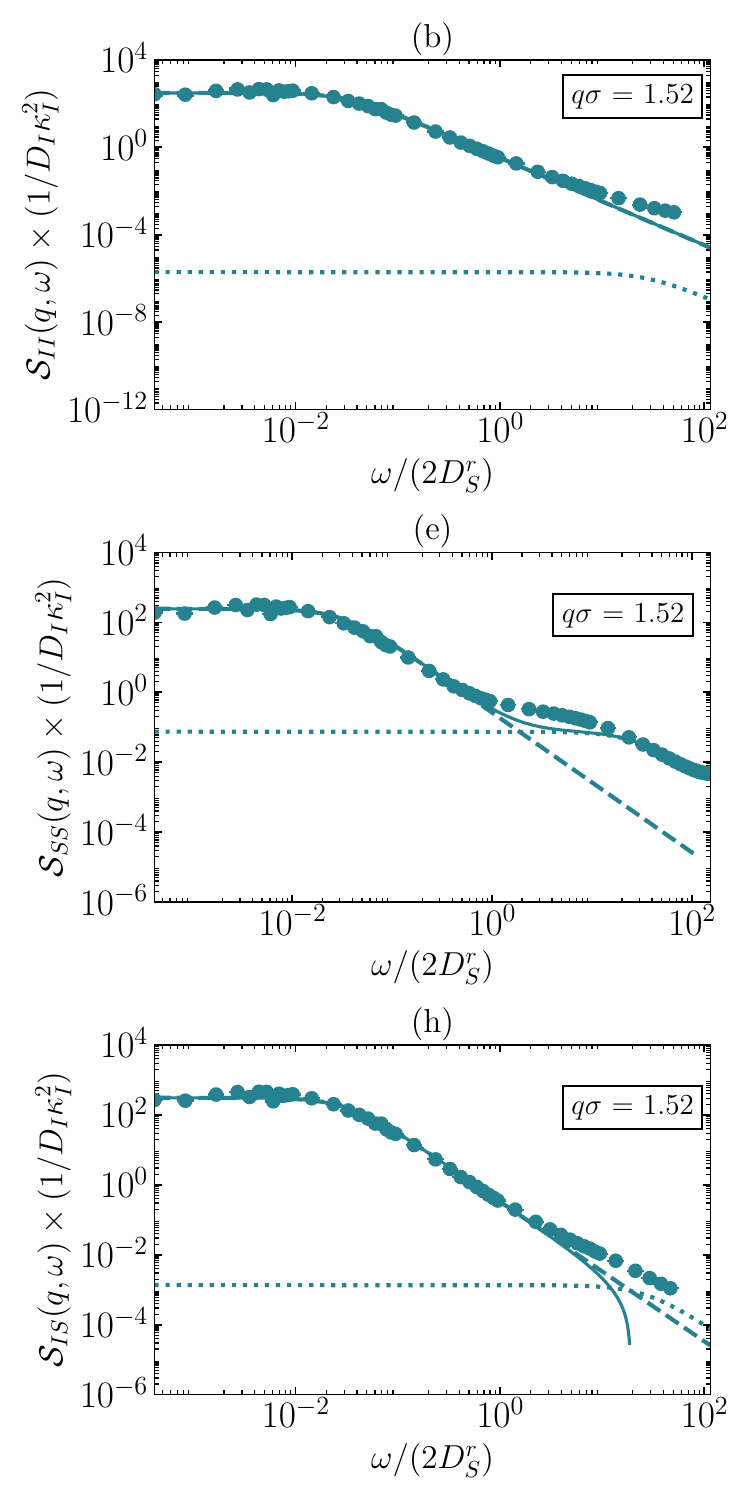} &
    \includegraphics[width=0.33\linewidth]{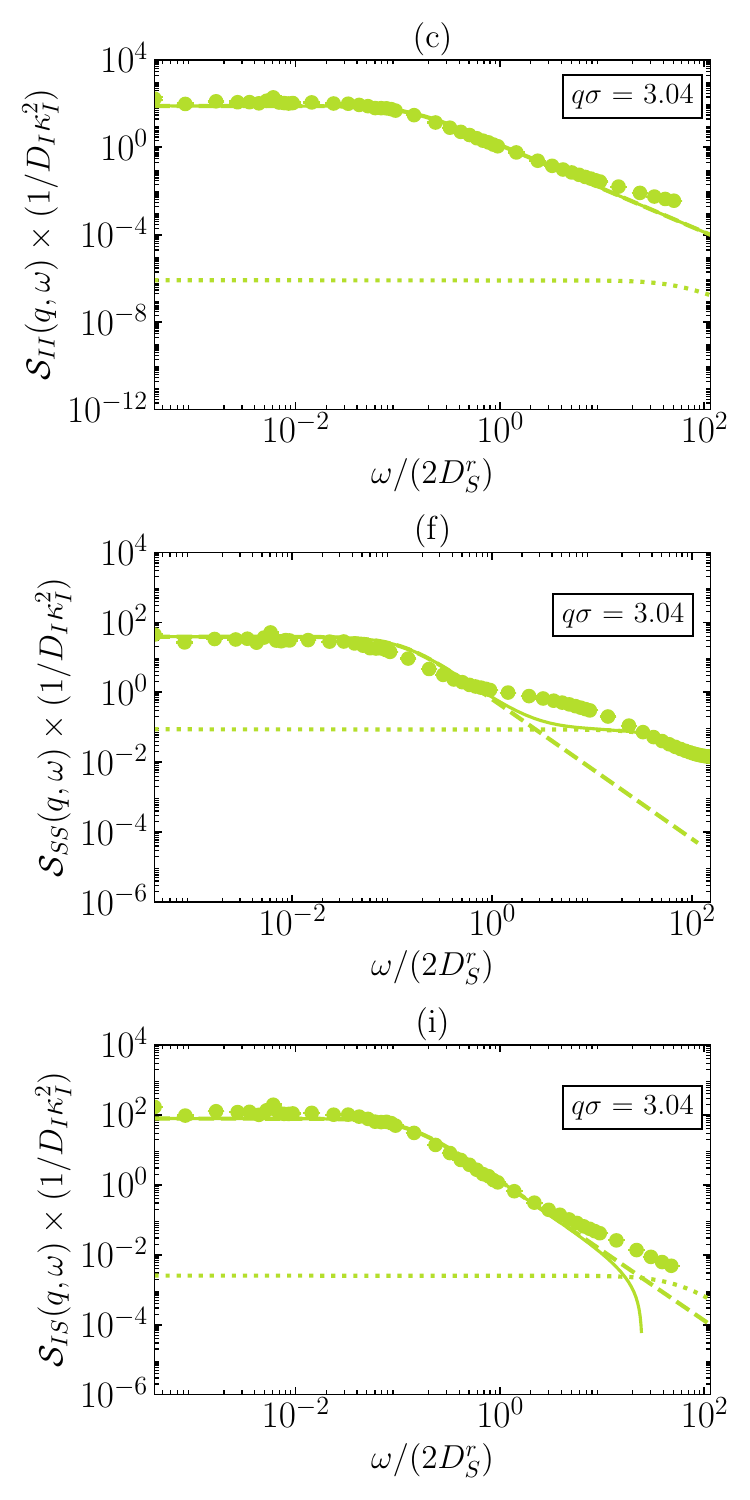} 
\end{tabular}
\caption{Dynamic charge structure factors $\mathcal{S}_{\alpha\beta}(\qq,\omega)$ at larger wavevectors for   System III, calculated from Brownian dynamics simulations (symbols) and SDFT (lines). Solid lines show the full SDFT predictions based on Eqs.~\eqref{eq_SII_qnonzero}-\eqref{eq_SS_qnonzero}, while dashed lines denote contribution from their second term only. In plots (a)-(f) dotted lines represent the contribution from the first term of Eqs.~\eqref{eq_SII_qnonzero} and \eqref{eq_SS_qnonzero}, whereas in plots (g)-(i) they correspond to the opposite of the first term of Eq.~\eqref{eq_SIS_q_nonzero}. 
 The errorbars denote one standard error.}
 \label{fig_DCSF_q_nonzero_SM_0_4M}   
\end{figure*}

This section compares SDFT predictions with BD results for the additional systems considered in this study. As explained in the previous section, we first apply the calibration methodology to obtain the optimal values for $\kappa_{I}$, $T_{S}$ and $T_{I}$.
Table~\ref{tab_SDFT_para} summarizes the calibrated parameters, while Figs.~\ref{fig_DCSF_q0_SM}-\ref{fig_DCSF_q_nonzero_SM_0_4M} present comparisons between SDFT predictions and BD results for the dynamic charge structure factors. As discussed in the main text, SDFT shows excellent agreement with BD across all systems, particularly in the low-frequency regime, even at large wavevectors. Figs.~\ref{fig_DCSF_q_nonzero_SM_p0_9} and \ref{fig_DCSF_q_nonzero_SM_0_4M} further demonstrate a clear quantitative separation between the regimes dominated by the two relaxation modes predicted by SDFT.
For $\mathcal{S}_{II}(\qq,\omega)$ and $\mathcal{S}_{IS}(\qq,\omega)$, solvent contributions are negligible, and the dynamics is dominated by the ionic relaxation mode. Whereas, for 
$\mathcal{S}_{SS}(\qq,\omega)$ in the case of fast-relaxing solvents, we can clearly see the distinct two step relaxation, with the high frequency regime governed by the solvent relaxation and the low frequency regime controlled by the ionic relaxation. 

\begin{table*}
    \centering
    \begin{tabular}{c|c|c|c|c|c|c}
         \textbf{System} & $\kappa^{\mathrm{bare}}_{I}\ \mathrm{(nm^{-1})}$ & $T^{\mathrm{bare}}_{S}\left(= \frac{1}{2D^{r}_{S}\varepsilon_{S}}\right) \mathrm{(ps)}$ & $T^{\mathrm{bare}}_{I}\left(= \frac{\varepsilon_{S}}{D_{I}\kappa_{I}^{2}}\right)\ \mathrm{(ps)}$ & $\kappa^{\mathrm{eff}}_{I}\ \mathrm{(nm^{-1})}$ & $T^{\mathrm{eff}}_{S}\ \mathrm{(ps)}$ & $T^{\mathrm{eff}}_{I}\ \mathrm{(ps)}$\\
         \hline 
         II & 9.50 &  0.156 & $1.39\times 10^{4}$ & 2.96 & 0.741 & $1.88\times10^{5}$ \\
         III & 18.4 & 0.0682 & $8.49\times 10^{3}$ & 10.0 & 0.565 & $1.14\times10^{4}$ \\ 
    \end{tabular}
    \caption{Values of parameters used in the SDFT predictions. For the `bare' values, we use $D^r_{S} = 0.05\ \mathrm{ps}^{-1}$ and set $D_{I}$ to the long-time ionic diffusion coefficient (see Table~\ref{tab_diffcoeff} for each system). Parameters marked with superscript `eff' indicate calibrated values used in computing the dynamic charge structure factors. For all cases (including System-I) $a = 0.214\ \mathrm{nm}$.}
    \label{tab_SDFT_para}
\end{table*}


\subsection{Comparison of calibrated parameters}

\begin{figure}
\begin{tabular}{cc}
    \includegraphics[width=0.3\linewidth]{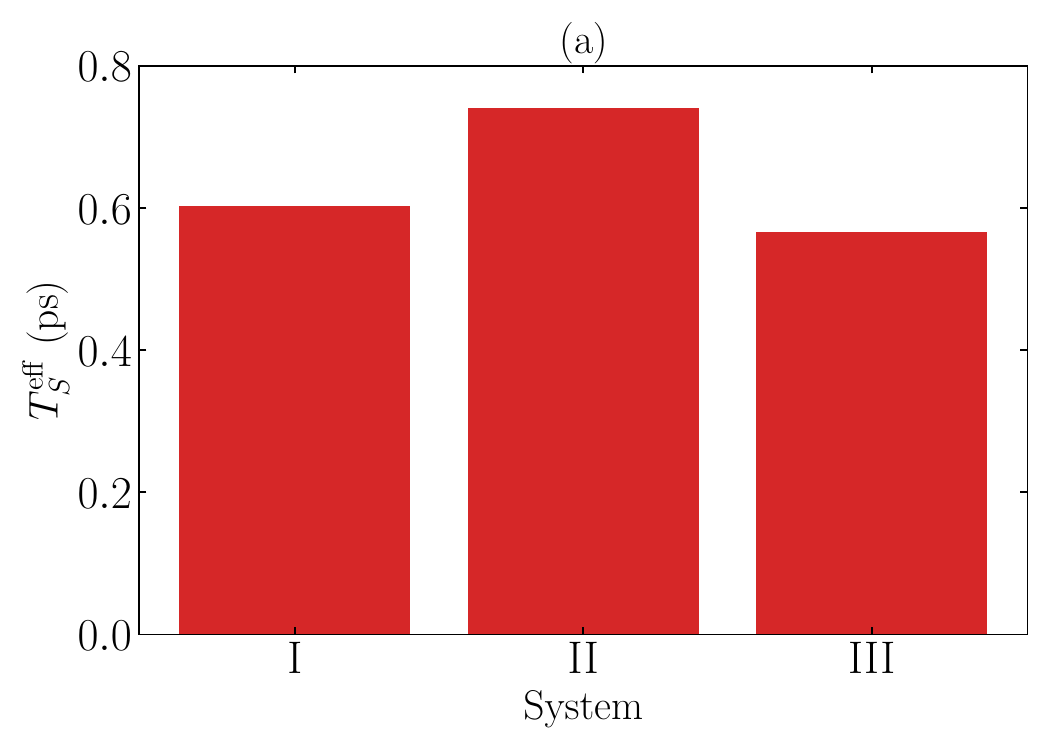} &
    \includegraphics[width=0.3\linewidth]{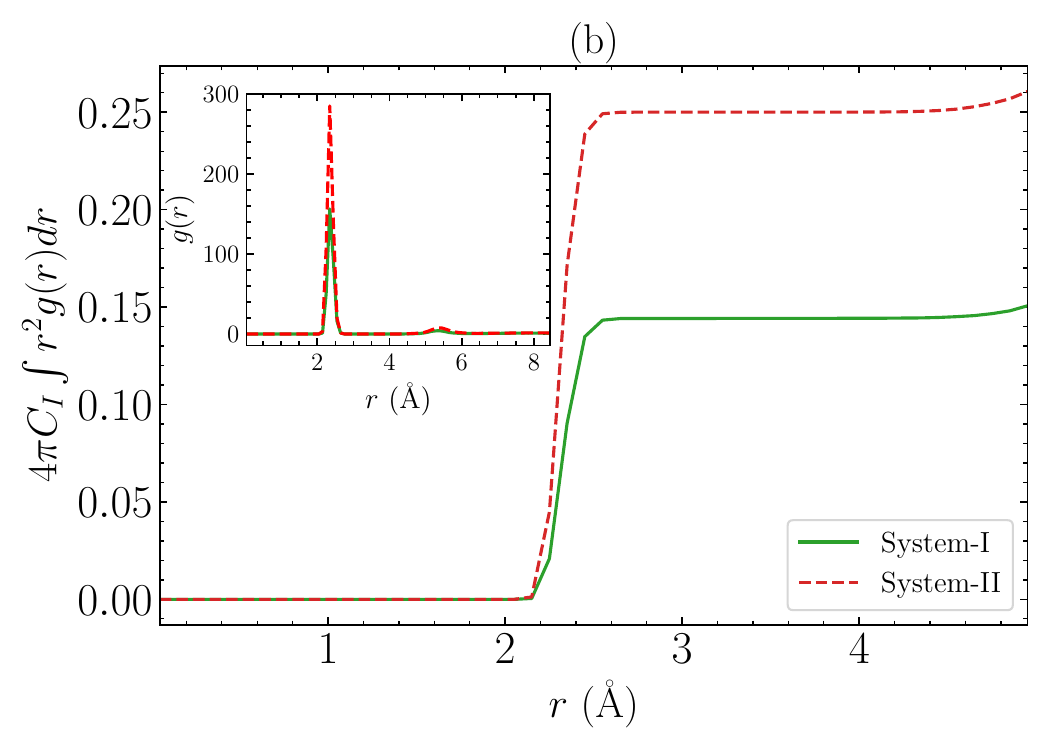} \\
    \includegraphics[width=0.3\linewidth]{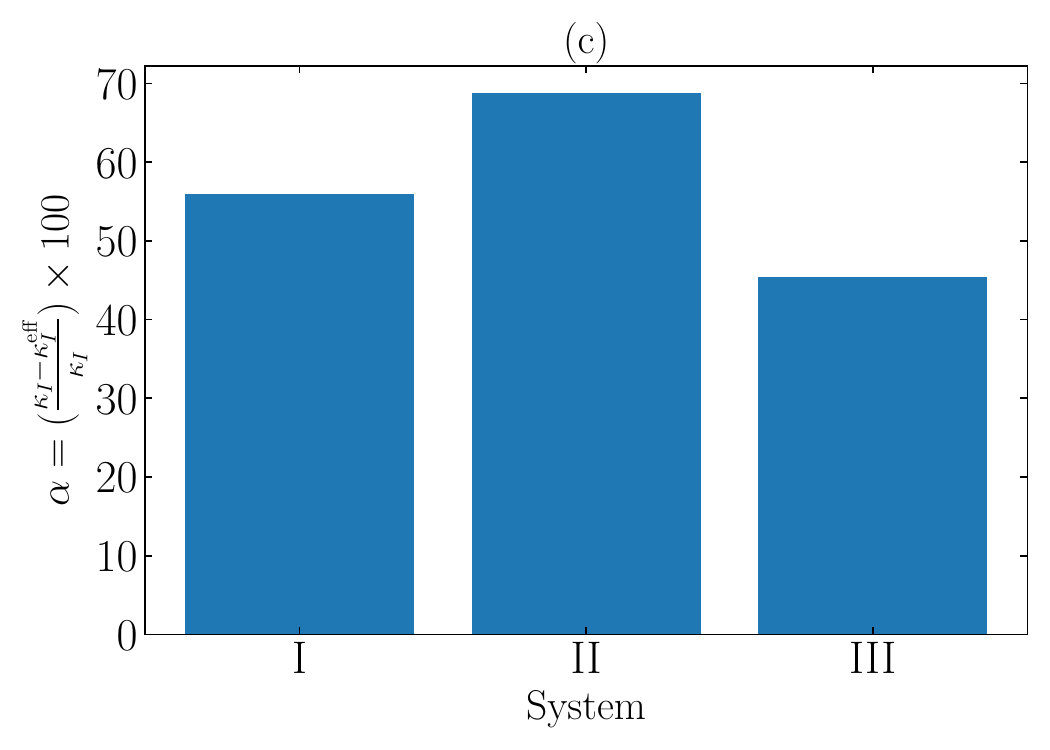}  &
    \includegraphics[width=0.3\linewidth]{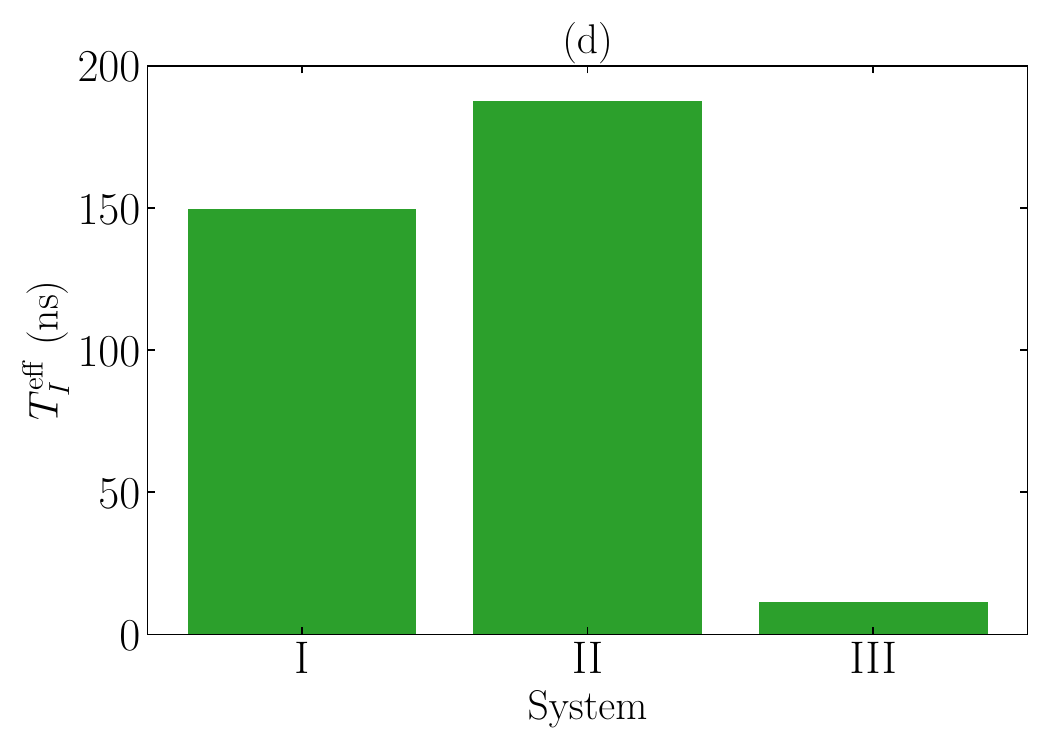} 
    
\end{tabular}
 \caption{(a), (c) and (d): Variation of calibrated parameters across studied systems. (b) Cation-anion coordination number for the 0.1 M systems. Inset plot in (b) is the radial distribution function between cation and anion.}
 \label{fig_params_compare}   
\end{figure}

Fig.~\ref{fig_params_compare} illustrates the variation of the calibrated parameters across the different electrolyte systems considered in this study. One key consequence of increasing solvent permittivity is the slowing down of solvent relaxation dynamics~\cite{Varghese2025}, which is clearly evident in Fig.~\ref{fig_params_compare}(a). Solvent permittivity also plays a crucial role in determining the extent of ion pairing within an electrolyte. Lower permittivity leads to weaker electrostatic screening, promoting the formation of ion pairs, also known as Bjerrum pairs. Fig~\ref{fig_params_compare}(b) shows the cation-anion coordination number for the 0.1 M systems. The non-zero coordination number within the first coordination sphere indicates the presence of ion-pairing. As mentioned above, the extent of ion-pairing is strongly influenced by the solvent permittivity, with lower permittivity solvents favoring greater ion association.
This reduces the concentration of free ions available for screening, resulting in an effective ionic concentration that is lower than that of a fully dissociated electrolyte~\cite{zwanikken2009inflation, adar2017bjerrum}. 
This behavior is reflected in the calibrated values of $\kappa_{I}$, since $\kappa_{I} \propto \sqrt{C_{I}}$. Fig.~\ref{fig_params_compare}(b) shows the percentage reduction in $\kappa_{I}$ relative to its nominal value, denoted by $\alpha$, for the different systems. As expected, the largest reduction ($\sim70\%$) occurs for the system with the lowest solvent permittivity ($\varepsilon_S = 64.07$), whereas systems with higher permittivity ($\varepsilon_S = 146.5$) exhibit more moderate reductions, with $\alpha \sim 40–50\%$. Finally, the ionic relaxation times, shown in Fig.~\ref{fig_params_compare}(c), indicate the fastest relaxation for the 0.4 M system, as anticipated, while the 0.1 M systems exhibit relatively similar values of $T^\text{eff}_{I}$.

\subsection{Translational diffusion coefficient}

\begin{figure}[ht]
\begin{tabular}{cc}
    \includegraphics[width=0.3\columnwidth]{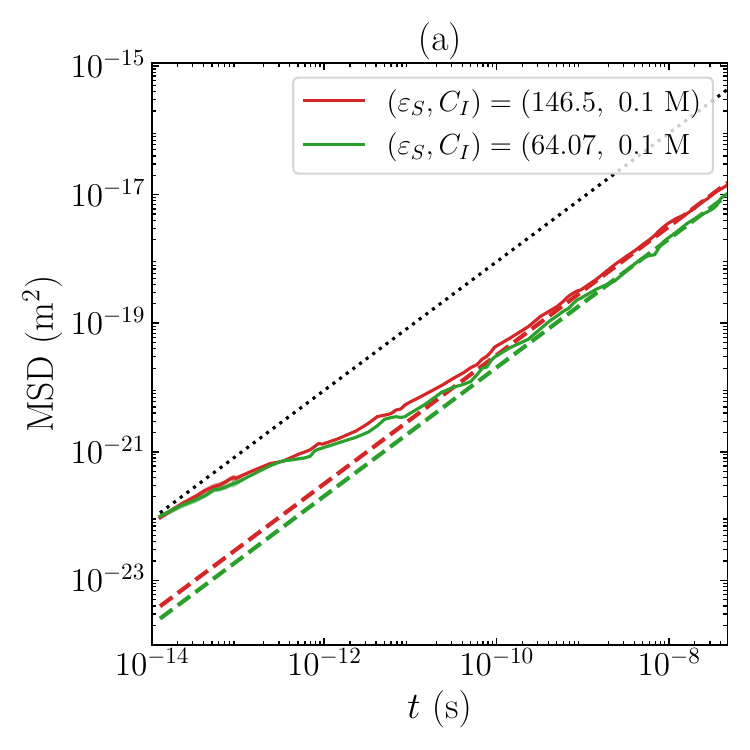} &
    \includegraphics[width=0.3\columnwidth]{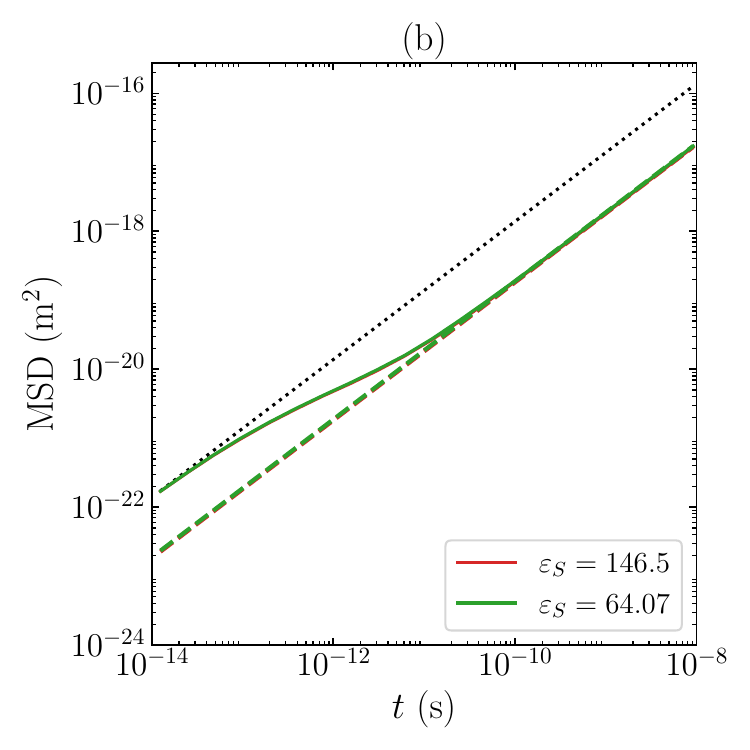} 
     
\end{tabular}
    \caption{ Mean square displacement plots for the (a) ion and (b) solvent particles. 
    Solid lines correspond to MSD data from BD simulations and dashed lines represent $6 D^{\mathrm{fit}}_{\alpha} t$, where $D^{\mathrm{fit}}_{\alpha}$ is the long-time diffusion coefficient and $\alpha \in \{I, S\}$. The dotted line denotes the MSD plot for the non-interacting particles limit. 
    Plots in (a) are based on electrolyte systems, while the plots in (b) are obtained from polar fluid systems. }
    \label{fig_MSD_SM}
\end{figure}

In this section, we examine the effect of interactions on the transport properties of an electrolyte system. The Fig.~\ref{fig_MSD_SM} shows the mean square displacement (MSD) of both ion and solvent particles undergoing Brownian motion. From MSD we can obtain the particle diffusion coefficient by fitting its long-time behavior to $6Dt$. At short times, both species follow their respective bare diffusion coefficients, followed by a caging regime at intermediate times, and a long-time diffusive plateau. The corresponding long-time diffusion coefficients, extracted from the MSD and summarized in Table~\ref{tab_diffcoeff}, are governed primarily by steric interactions, with electrostatic contributions remaining negligible. We further find that the presence of ions does not measurably affect the translational diffusion of the solvent, in a 0.1 M electrolyte system (Table~\ref{tab_diffcoeff}). 
Importantly, Fig.~\ref{fig_MSD_SM}(a) offers a corrected estimate for the ionic diffusion coefficient $D_I$ used in calibrating the SDFT model, as the present SDFT framework requires the steady-state value of this parameter.

\begin{table*}
    \centering
    \begin{tabular}{l|c|c}
         \textbf{System} ($\varepsilon_{S}, C_{I}$) & \textbf{Ion diffusion coefficient 
         ($\mathrm{m}^{2}\mathrm{s}^{-1}$)} & \textbf{Solvent diffusion coefficient 
         ($\mathrm{m}^{2}\mathrm{s}^{-1}$)} \\
         \hline 
         Electrolyte (146.5, 0.1 M) & $5.10\times10^{-11}$ & $2.96\times10^{-10}$ \\
         Electrolyte (64.07, 0.1 M) & $3.45\times10^{-11}$ & $3.04\times10^{-10}$ \\
         Polar fluid (146.5) & - &  $3.12\times10^{-10}$\\ 
         Polar fluid (64.07) & - & $3.01\times10^{-10}$
    \end{tabular}
    \caption{Long-time diffusion coefficients calculated from the MSD. The corresponding bare values for the ion and solvent translation diffusion coefficients are $D_{I} = 1.5 \times 10^{-9} \ \mathrm{m}^{2}\ \mathrm{s}^{-1}$ and $D_{S} = 2.3 \times 10^{-9} \ \mathrm{m}^{2}\ \mathrm{s}^{-1}$, respectively. 
    }
    \label{tab_diffcoeff}
\end{table*}

\section{Beyond the limit of a fast-relaxing solvent}\label{app_slow}

From the general expressions in {Appendix~\ref{sec_DCSF_general}}, we deduce the observables that are discussed in the main text.  First, the ionic relaxation time is defined as
$\tau_{I}^\text{relax }(q) = [\int_{0}^{\infty}  F_{II}(q,t) \, \dd t]/S_{II}(q)$, for any $q$. In dimensionless variables, we get
\begin{equation}
\tau_{I}^\text{relax }(Q)= \frac{K_I^2 T_I \left(Q^2 + 2\zeta \epsS - 2\zeta + 2\right)}
{(Q^2 + 2)\left(K_I^2 + Q^2 \epsS\right)}
\end{equation}
For $Q=0$, we retrieve the expression given in the main text. Second, the ratio between the two eigenvalues $\lambda_{\pm 1}$ reads 
    \begin{equation}
    \frac{\lambda_{-1}(Q)}{\lambda_1(Q)}=\frac{
(Q^2 + 2\zeta \epsS + 2)K_I^2 + 2Q^2 \zeta \epsS - \Delta
}{
(Q^2 + 2\zeta \epsS + 2)K_I^2 + 2Q^2 \zeta \epsS + \Delta
}.
\end{equation}
Finally, the ratio between the two weights $\mathcal{F}_{II}^{(\pm1)}$ reads
\begin{equation}
    \frac{\mathcal{F}_{II}^{(-1)}(Q)}{\mathcal{F}_{II}^{(1)}(Q)}
    =
   -\frac{
\left(
\Delta + \left((2\epsS-4)\zeta + Q^2 + 2\right)K_I^2
-2Q^2\zeta\epsS
\right)^2
}{
-16K_I^2Q^2\zeta^2\epsS^2
-16K_I^4\zeta^2\epsS
+16K_I^2Q^2\zeta^2\epsS
+16K_I^4\zeta^2
}
\end{equation}
which yields, in the limit $Q\to 0$:
\begin{equation}
 \frac{\mathcal{F}_{II}^{(-1)}(0)}{\mathcal{F}_{II}^{(1)}(0)}
 =
\frac{
\left(
\zeta\epsS
+\sqrt{\zeta^2\epsS^2 + 2\zeta\epsS - 4\zeta + 1}
-2\zeta + 1
\right)^2
}{
4\zeta^2(\epsS - 1)
}
\end{equation}

\twocolumngrid

\bibliographystyle{apsrev4-1}
\bibliography{biblio2.bib}

@article{Illien2024e,
  title = {Stochastic {{Density Functional Theory}} for {{Ions}} in a {{Polar Solvent}}},
  author = {Illien, Pierre and Carof, Antoine and Rotenberg, Benjamin},
  year = 2024,
  month = dec,
  journal = {Phys. Rev. Lett.},
  volume = {133},
  number = {26},
  pages = {268002},
  issn = {0031-9007, 1079-7114},
  doi = {10.1103/PhysRevLett.133.268002},
  urldate = {2025-01-16}
}

@article{Chandra1999,
  title = {Ion Conductance in Electrolyte Solutions},
  author = {Chandra, Amalendu and Bagchi, Biman},
  year = 1999,
  journal = {J. Chem. Phys.},
  volume = {110},
  number = {20},
  pages = {10024--10034},
  issn = {00219606},
  doi = {10.1063/1.478876}
}

@article{Chandra2000a,
  title = {Frequency Dependence of Ionic Conductivity of Electrolyte Solutions},
  author = {Chandra, Amalendu and Bagchi, Biman},
  year = 2000,
  journal = {J. Chem. Phys.},
  volume = {112},
  pages = {1876},
  issn = {00219606},
  doi = {10.1063/1.480751}
}

@article{Dufreche2002a,
  title = {Ionic {{Self-Diffusion}} in {{Concentrated Aqueous Electrolyte Solutions}}},
  author = {Dufr{\^e}che, J.-F. and Bernard, O. and Turq, P. and Mukherjee, A. and Bagchi, B.},
  year = 2002,
  month = feb,
  journal = {Phys. Rev. Lett.},
  volume = {88},
  number = {9},
  pages = {095902},
  issn = {0031-9007, 1079-7114},
  doi = {10.1103/PhysRevLett.88.095902},
  urldate = {2024-07-07}
}

@article{Dufreche2005a,
  title = {Transport in Electrolyte Solutions: Are Ions {{Brownian}} Particles?},
  shorttitle = {Transport in Electrolyte Solutions},
  author = {Dufr{\^e}che, J.-F. and Bernard, O. and Turq, P.},
  year = 2005,
  month = apr,
  journal = {Journal of Molecular Liquids},
  volume = {118},
  number = {1-3},
  pages = {189--194},
  issn = {01677322},
  doi = {10.1016/j.molliq.2004.07.036},
  urldate = {2026-03-18}
}

@article{Roy2015,
  title = {Mode Coupling Theory Analysis of Electrolyte Solutions: Time Dependent Diffusion, Intermediate Scattering Function, and Ion Solvation Dynamics},
  author = {Roy, Susmita and Yashonath, Subramanian and Bagchi, Biman},
  year = 2015,
  journal = {J. Chem. Phys.},
  volume = {142},
  pages = {124502},
  number = {12},
  doi = {10.1063/1.4915274}
}

@article{Bagchi1998,
  title = {Ionic {{Mobility}} and {{Ultrafast Solvation}}: {{Control}} of a {{Slow Phenomenon}} by {{Fast Dynamics}}},
  shorttitle = {Ionic {{Mobility}} and {{Ultrafast Solvation}}},
  author = {Bagchi, Biman and Biswas, Ranjit},
  year = 1998,
  month = apr,
  journal = {Acc. Chem. Res.},
  volume = {31},
  number = {4},
  pages = {181--187},
  issn = {0001-4842, 1520-4898},
  doi = {10.1021/ar970226f},
  urldate = {2026-02-24}
}

@article{Buchner2009,
  title = {Interactions and Dynamics in Electrolyte Solutions by Dielectric Spectroscopy},
  author = {Buchner, Richard and Hefter, Glenn},
  year = 2009,
  journal = {Phys. Chem. Chem. Phys.},
  volume = {11},
  number = {40},
  pages = {8984},
  issn = {1463-9076, 1463-9084},
  doi = {10.1039/b906555p},
  urldate = {2026-03-02}
}

@article{Yamaguchi2007,
  title = {A Theoretical Study on the Frequency-Dependent Electric Conductivity of Electrolyte Solutions},
  author = {Yamaguchi, T. and Matsuoka, T. and Koda, S.},
  year = 2007,
  journal = {J. Chem. Phys.},
  volume = {127},
  pages = {234501},
  issn = {00219606},
  doi = {10.1063/1.2806289}
}

@article{Bonneau2024,
  title = {Frequency-Dependent Conductivity of Concentrated Electrolytes: {{A}} Stochastic Density Functional Theory},
  shorttitle = {Frequency-Dependent Conductivity of Concentrated Electrolytes},
  author = {Bonneau, Haggai and Avni, Yael and Andelman, David and Orland, Henri},
  year = 2024,
  month = dec,
  journal = {J. Chem. Phys.},
  volume = {161},
  number = {24},
  pages = {244501},
  issn = {0021-9606, 1089-7690},
  doi = {10.1063/5.0236073},
  urldate = {2026-03-03}
}

@article{Donev2019,
  title = {Fluctuating {{Hydrodynamics}} and {{Debye-H\"uckel-Onsager Theory}} for {{Electrolytes}}},
  author = {Donev, Aleksandar and Garcia, Alejandro L. and P{\'e}raud, Jean Philippe and Nonaka, Andrew J. and Bell, John B.},
  year = 2019,
  journal = {Curr. Opin. Electrochem.},
  volume = {13},
  pages = {1--10},
  publisher = {Elsevier Ltd},
  issn = {24519111},
  doi = {10.1016/j.coelec.2018.09.004}
}

@article{Peraud2017,
  title = {Fluctuation-Enhanced Electric Conductivity in Electrolyte Solutions},
  author = {P{\'e}raud, Jean Philippe and Nonaka, Andrew J. and Bell, John B. and Donev, Aleksandar and Garcia, Alejandro L.},
  year = 2017,
  journal = {Proc. Natl. Acad. Sci. U.S.A.},
  volume = {114},
  number = {41},
  pages = {10829--10833},
  issn = {10916490},
  doi = {10.1073/pnas.1714464114}
}

@article{Illien2025a,
  title = {The {{Dean}}--{{Kawasaki}} Equation and Stochastic Density Functional Theory},
  author = {Illien, Pierre},
  year = 2025,
  month = aug,
  journal = {Rep. Prog. Phys.},
  volume = {88},
  number = {8},
  pages = {086601},
  issn = {0034-4885, 1361-6633},
  doi = {10.1088/1361-6633/adee2e},
  urldate = {2025-09-17}
}

@article{Dean1996,
  title = {Langevin Equation for the Density of a System of Interacting {{Langevin}} Processes},
  author = {Dean, D S},
  year = 1996,
  journal = {J. Phys. A: Math. Gen.},
  volume = {29},
  pages = {L613},
  issn = {0305-4470},
  doi = {10.1088/0305-4470/29/24/001}
}

@article{Kawasaki1994,
  title = {Stochastic Model of Slow Dynamics in Supercooled Liquids and Dense Colloidal Suspensions},
  author = {Kawasaki, Kyozi},
  year = 1994,
  journal = {Physica A},
  volume = {208},
  pages = {35},
  issn = {03784371},
  doi = {10.1016/0378-4371(94)90533-9},
  isbn = {03784371}
}

@article{Varghese2025,
  title = {Dynamic Correlations in a Polar Fluid: {{Confronting}} Stochastic Density Functional Theory to Simulations},
  shorttitle = {Dynamic Correlations in a Polar Fluid},
  author = {Varghese, Sleeba and Illien, Pierre and Rotenberg, Benjamin},
  year = 2025,
  month = sep,
  journal = {J. Chem. Phys.},
  volume = {163},
  number = {12},
  pages = {124107},
  issn = {0021-9606, 1089-7690},
  doi = {10.1063/5.0292306},
  urldate = {2025-10-08}
}

@article{Bonneau2023,
  title = {Temporal Response of the Conductivity of Electrolytes},
  author = {Bonneau, Haggai and D{\'e}mery, Vincent and Rapha{\"e}l, {\'E}lie},
  year = 2023,
  journal = {J. Stat. Mech.},
  volume = {2023},
  pages = {073205},
  doi = {10.1088/1742-5468/acdced}
}

@article{Demery2016,
  title = {The Conductivity of Strong Electrolytes from Stochastic Density Functional Theory},
  author = {D{\'e}mery, Vincent and Dean, David S},
  year = 2016,
  journal = {J. Stat. Mech.},
  volume = {2016},
  pages = {023106},
  issn = {17425468},
  doi = {10.1088/1742-5468/2016/02/023106}
}

@article{Avni2022,
  title = {Conductance of Concentrated Electrolytes: {{Multivalency}} and the {{Wien}} Effect},
  shorttitle = {Conductance of Concentrated Electrolytes},
  author = {Avni, Yael and Andelman, David and Orland, Henri},
  year = 2022,
  month = oct,
  journal = {J. Chem. Phys.},
  volume = {157},
  number = {15},
  pages = {154502},
  doi = {10.1063/5.0111645},
  urldate = {2024-07-07},
}

@article{Avni2022a,
  title = {Conductivity of {{Concentrated Electrolytes}}},
  author = {Avni, Yael and Adar, Ram M. and Andelman, David and Orland, Henri},
  year = 2022,
  month = mar,
  journal = {Phys. Rev. Lett.},
  volume = {128},
  number = {9},
  pages = {098002},
  issn = {0031-9007, 1079-7114},
  doi = {10.1103/PhysRevLett.128.098002},
  urldate = {2024-07-07}
}

@article{Bernard2023a,
  title = {On Analytical Theories for Conductivity and Self-Diffusion in Concentrated Electrolytes},
  author = {Bernard, Olivier and Jardat, Marie and Rotenberg, Benjamin and Illien, Pierre},
  year = 2023,
  month = oct,
  journal = {J. Chem. Phys.},
  volume = {159},
  number = {16},
  pages = {164105},
  issn = {0021-9606, 1089-7690},
  doi = {10.1063/5.0165533},
  urldate = {2024-10-17}
}

@article{Robin2024,
  title = {Correlation-Induced Viscous Dissipation in Concentrated Electrolytes},
  author = {Robin, Paul},
  year = 2024,
  journal = {J. Chem. Phys.},
  volume = {160},
  pages = {064503},
  publisher = {AIP Publishing, LLC},
  issn = {10897690},
  doi = {10.1063/5.0188215},
  pmid = {38349632}
}

@article{Bonneau2025,
  title = {Stationary and Transient Correlations in Driven Electrolytes},
  author = {Bonneau, Haggai and D{\'e}mery, Vincent and Rapha{\"e}l, Elie},
  year = 2025,
  month = mar,
  journal = {J. Stat. Mech.},
  volume = {2025},
  number = {3},
  pages = {033201},
  issn = {1742-5468},
  doi = {10.1088/1742-5468/adb4ce},
  urldate = {2025-05-08}
}

@article{Mahdisoltani2021a,
  title = {Long-{{Range Fluctuation-Induced Forces}} in {{Driven Electrolytes}}},
  author = {Mahdisoltani, Saeed and Golestanian, Ramin},
  year = 2021,
  month = apr,
  journal = {Phys. Rev. Lett.},
  volume = {126},
  number = {15},
  pages = {158002},
  issn = {0031-9007, 1079-7114},
  doi = {10.1103/PhysRevLett.126.158002},
  urldate = {2024-07-07}
}

@article{Mahdisoltani2021c,
  title = {Transient Fluctuation-Induced Forces in Driven Electrolytes after an Electric Field Quench},
  author = {Mahdisoltani, Saeed and Golestanian, Ramin},
  year = 2021,
  journal = {New J. Phys.},
  volume = {23},
  pages = {073034},
  publisher = {IOP Publishing},
  issn = {13672630},
  doi = {10.1088/1367-2630/ac0f1a}
}

@article{Dean2014a,
  title = {Relaxation of the Thermal {{Casimir}} Force between Net Neutral Plates Containing {{Brownian}} Charges},
  author = {Dean, David S. and Podgornik, Rudolf},
  year = 2014,
  month = mar,
  journal = {Phys. Rev. E},
  volume = {89},
  pages = {032117},
  issn = {1539-3755, 1550-2376},
  doi = {10.1103/PhysRevE.89.032117},
  urldate = {2024-07-07}
}

@article{Du2024,
  title = {Correlation {{Decoupling}} of {{Casimir Interaction}} in an {{Electrolyte Driven}} by {{External Electric Fields}}},
  author = {Du, Guangle and Dean, David S. and Miao, Bing and Podgornik, Rudolf},
  year = 2024,
  month = dec,
  journal = {Phys. Rev. Lett.},
  volume = {133},
  number = {23},
  pages = {238002},
  issn = {0031-9007, 1079-7114},
  doi = {10.1103/PhysRevLett.133.238002},
  urldate = {2024-12-09}
}

@article{Du2025,
  title = {Repulsive Thermal van Der {{Waals}} Interaction in Multispecies Asymmetric Electrolytes Driven by External Electric Fields},
  author = {Du, Guangle and Dean, David S. and Miao, Bing and Podgornik, Rudolf},
  year = 2025,
  month = apr,
  journal = {Phys. Rev. E},
  volume = {111},
  number = {4},
  pages = {044108},
  issn = {2470-0045, 2470-0053},
  doi = {10.1103/PhysRevE.111.044108},
  urldate = {2025-05-08}
}

@article{Lu2015,
  title = {Out-of-Equilibrium Thermal {{Casimir}} Effect between {{Brownian}} Conducting Plates},
  author = {Lu, B.-S. and Dean, D. S. and Podgornik, R.},
  year = 2015,
  month = oct,
  journal = {EPL},
  volume = {112},
  number = {2},
  pages = {20001},
  issn = {0295-5075, 1286-4854},
  doi = {10.1209/0295-5075/112/20001},
  urldate = {2024-07-07}
}

@article{Bagchi1989,
  title = {Polarization Relaxation, Dielectric Dispersion, and Solvation Dynamics in Dense Dipolar Liquid},
  author = {Bagchi, Biman and Chandra, Amalendu},
  year = 1989,
  month = jun,
  journal = {J. Chem. Phys.},
  volume = {90},
  number = {12},
  pages = {7338--7345},
  issn = {0021-9606, 1089-7690},
  doi = {10.1063/1.456213},
  urldate = {2025-04-24}
}

@article{Chandra1988,
  title = {The Role of Translational Diffusion in the Polarization Relaxation in Dense Polar Liquids},
  author = {Chandra, A and Bagchi, B},
  year = 1988,
  journal = {Chem. Phys. Lett.},
  volume = {151},
  pages = {47},
  doi = {10.1016/0009-2614(88)80067-8}
}

@article{Vijayadamodar1989,
  title = {Effects of Translational Diffusion on Dielectric Friction in a Dipolar Liquid},
  author = {Vijayadamodar, G.V. and Chandra, A. and Bagchi, B.},
  year = 1989,
  month = sep,
  journal = {Chem. Phys. Lett.},
  volume = {161},
  number = {4-5},
  pages = {413--419},
  issn = {00092614},
  doi = {10.1016/0009-2614(89)85108-5},
  urldate = {2025-06-02}
}

@article{Chandra1990a,
  title = {Relationship between Microscopic and Macroscopic Orientational Relaxation Times in Polar Liquids},
  author = {Chandra, {\relax Amalendu}. and Bagchi, {\relax Biman}.},
  year = 1990,
  month = apr,
  journal = {J. Phys. Chem.},
  volume = {94},
  number = {7},
  pages = {3152--3156},
  issn = {0022-3654, 1541-5740},
  doi = {10.1021/j100370a074},
  urldate = {2025-06-11}
}

@article{Thompson2022,
  author = {Thompson, Aidan P. and Aktulga, H. Metin and Berger, Richard and Bolintineanu, Dan S. and Brown, W. Michael and Crozier, Paul S. and {in 't Veld}, Pieter J. and Kohlmeyer, Axel and Moore, Stan G. and Nguyen, Trung Dac and Shan, Ray and Stevens, Mark J. and Tranchida, Julien and Trott, Christian and Plimpton, Steven J.},
  journal = {Comput. Phys. Commun.},
  pages = {108171},
  publisher = {Elsevier B.V.},
  title = {{{LAMMPS}} - a Flexible Simulation Tool for Particle-Based Materials Modeling at the Atomic, Meso, and Continuum Scales},
  volume = {271},
  year = {2022},
  doi = {10.1016/j.cpc.2021.108171},
  issn = {00104655},
}

@article{Zorkot2016,
  title = {The {{Power Spectrum}} of {{Ionic Nanopore Currents}}: {{The Role}} of {{Ion Correlations}}},
  shorttitle = {The {{Power Spectrum}} of {{Ionic Nanopore Currents}}},
  author = {Zorkot, Mira and Golestanian, Ramin and Bonthuis, Douwe Jan},
  year = 2016,
  month = apr,
  journal = {Nano Lett.},
  volume = {16},
  number = {4},
  pages = {2205--2212},
  issn = {1530-6984, 1530-6992},
  doi = {10.1021/acs.nanolett.5b04372},
  urldate = {2026-03-10}
}

@article{Zorkot2016a,
  title = {Current Fluctuations in Nanopores: {{The}} Effects of Electrostatic and Hydrodynamic Interactions},
  shorttitle = {Current Fluctuations in Nanopores},
  author = {Zorkot, Mira and Golestanian, Ramin and Bonthuis, Douwe Jan},
  year = 2016,
  month = oct,
  journal = {Eur. Phys. J. Spec. Top.},
  volume = {225},
  number = {8-9},
  pages = {1583--1594},
  issn = {1951-6355, 1951-6401},
  doi = {10.1140/epjst/e2016-60152-y},
  urldate = {2026-03-10}
}

@article{Zorkot2018,
  title = {Current Fluctuations across a Nano-Pore},
  author = {Zorkot, Mira and Golestanian, Ramin},
  year = 2018,
  month = apr,
  journal = {J. Phys.: Condens. Matter},
  volume = {30},
  number = {13},
  pages = {134001},
  issn = {0953-8984, 1361-648X},
  doi = {10.1088/1361-648X/aab016},
  urldate = {2026-03-10}
}

@article{toukmaji2000efficient,
  title = {Efficient particle-mesh {E}wald based approach to fixed and induced dipolar interactions},
  author = {Toukmaji, Abdulnour and Sagui, Celeste and Board, John and Darden, Tom},
  journal = {J. Chem. Phys.},
  volume = {113},
  number = {24},
  pages = {10913--10927},
  year = {2000},
  doi = {10.1063/1.1324708},
  publisher = {American Institute of Physics}
}

@article{Delong2015,
  author = {Delong, Steven and Usabiaga, Florencio Balboa and Donev, Aleksandar},
  journal = {J. Chem. Phys.},
  pages = {144107},
  title = {Brownian Dynamics of Confined Rigid Bodies},
  volume = {143},
  year = {2015},
  doi = {10.1063/1.4932062}
}

@article{Ilie2015,
  author = {Ilie, Ioana M. and Briels, Wim J. and Den Otter, Wouter K.},
  journal = {J. Chem. Phys.},
  month = mar,
  number = {11},
  pages = {114103},
  title = {An Elementary Singularity-Free {{Rotational Brownian Dynamics}} Algorithm for Anisotropic Particles},
  volume = {142},
  year = {2015},
  doi = {10.1063/1.4914322},
  issn = {0021-9606, 1089-7690},
}

@article{Duboisset2018,
  title = {Salt-Induced {{Long-to-Short Range Orientational Transition}} in {{Water}}},
  author = {Duboisset, Julien and Brevet, Pierre-Fran{\c c}ois},
  year = 2018,
  month = jun,
  journal = {Phys. Rev. Lett.},
  volume = {120},
  number = {26},
  pages = {263001},
  doi = {10.1103/physrevlett.120.263001},
  urldate = {2025-07-22}
}

@article{Duboisset2020,
  title = {Long-{{Range Orientational Organization}} of {{Dipolar}} and {{Steric Liquids}}},
  author = {Duboisset, Julien and Rondepierre, Fabien and Brevet, Pierre-Fran{\c c}ois},
  year = 2020,
  month = nov,
  journal = {J. Phys. Chem. Lett.},
  volume = {11},
  number = {22},
  pages = {9869--9875},
  issn = {1948-7185, 1948-7185},
  doi = {10.1021/acs.jpclett.0c02705},
  urldate = {2025-08-13}
}

@article{Belloni2018,
  title = {Screened {{Coulombic Orientational Correlations}} in {{Dilute Aqueous Electrolytes}}},
  author = {Belloni, Luc and Borgis, Daniel and Levesque, Maximilien},
  year = 2018,
  month = apr,
  journal = {J. Phys. Chem. Lett.},
  volume = {9},
  number = {8},
  pages = {1985--1989},
  publisher = {American Chemical Society (ACS)},
  issn = {1948-7185, 1948-7185},
  doi = {10.1021/acs.jpclett.8b00606},
  urldate = {2025-07-22}
}

@article{Borgis2018,
  title = {What {{Does Second-Harmonic Scattering Measure}} in {{Diluted Electrolytes}}?},
  author = {Borgis, Daniel and Belloni, Luc and Levesque, Maximilien},
  year = 2018,
  month = jul,
  journal = {J. Phys. Chem. Lett.},
  volume = {9},
  number = {13},
  pages = {3698--3702},
  publisher = {American Chemical Society},
  doi = {10.1021/acs.jpclett.8b01690},
  urldate = {2024-09-23}
}

@article{ramirez2002density,
  title = {Density functional theory of solvation in a polar solvent: Extracting the functional from homogeneous solvent simulations},
  author = {Ramirez, Rosa and Gebauer, Ralph and Mareschal, Michel and Borgis, Daniel},
  journal = {Phys. Rev. E},
  volume = {66},
  number = {3},
  pages = {031206},
  year = {2002},
  doi = {10.1103/PhysRevE.66.031206},
  publisher = {APS}
}

@article{Hubbard1977a,
  title = {Dielectric Dispersion and Dielectric Friction in Electrolyte Solutions. {{I}}.},
  author = {Hubbard, J. and Onsager, L.},
  year = 1977,
  journal = {J. Chem. Phys.},
  volume = {67},
  pages = {4850},
  doi={10.1063/1.434664}
}

@article{Becker2025,
  title = {Dielectric {{Properties}} of {{Aqueous Electrolytes}} at the {{Nanoscale}}},
  author = {Becker, Maximilian R. and Netz, Roland R. and Loche, Philip and Bonthuis, Douwe Jan and Mouhanna, Dominique and Berthoumieux, H{\'e}l{\`e}ne},
  year = 2025,
  month = apr,
  journal = {Phys. Rev. Lett.},
  volume = {134},
  number = {15},
  pages = {158001},
  issn = {0031-9007, 1079-7114},
  doi = {10.1103/PhysRevLett.134.158001},
  urldate = {2026-01-05}
}

@article{Demery2026,
  title = {Effect of Solvent Structure on the {{Wien}} Effect and Ionic Correlations at the Nanoscale},
  author = {D{\'e}mery, Vincent and Toquer, Damien and Berthoumieux, H{\'e}l{\`e}ne},
  year = 2026,
  journal = {Faraday Discuss.},
  pages = {10.1039.D5FD00149H},
  issn = {1359-6640, 1364-5498},
  doi = {10.1039/D5FD00149H},
  urldate = {2026-03-09}
}

@misc{Maity2025,
  title = {What {{Dictates}} the {{Optimal Concentration}} of {{Li-Based Battery Electrolytes}}? {{A Combined Experimental}} and {{Simulation Study}}},
  shorttitle = {What {{Dictates}} the {{Optimal Concentration}} of {{Li-Based Battery Electrolytes}}?},
  author = {Maity, Narayan Chandra and Mitra, Sudipta and Biswas, Ranjit},
  year = 2025,
  month = may,
  doi = {10.26434/chemrxiv-2025-m4c68},
  urldate = {2026-03-17}
}

@article{Sha2019,
  title = {Dynamical Properties of a Room Temperature Ionic Liquid: {{Using}} Molecular Dynamics Simulations to Implement a Dynamic Ion Cage Model},
  shorttitle = {Dynamical Properties of a Room Temperature Ionic Liquid},
  author = {Sha, Maolin and Ma, Xiaohang and Li, Na and Luo, Fabao and Zhu, Guanglai and Fayer, Michael D.},
  year = 2019,
  month = oct,
  journal = {J. Chem. Phys.},
  volume = {151},
  number = {15},
  pages = {154502},
  issn = {0021-9606, 1089-7690},
  doi = {10.1063/1.5126231},
  urldate = {2026-03-17}
}

@article{Nakamoto2025,
  title = {Dielectric Behavior of Propylene Carbonate Solutions with {{LiPF6}}, {{LiClO4}}, and {{LiBF4}}. {{II}}. {{Molecular}} Dynamics Simulation Study},
  author = {Nakamoto, Mitsunori and Endo, Kota and Katayama, Shinichi and Han, Jihae and Otani, Erika and Watanabe, Hikari and Okae, Izaya and Umebayashi, Yasuhiro},
  year = 2025,
  month = nov,
  journal = {J. Chem. Phys.},
  volume = {163},
  number = {20},
  pages = {204504},
  issn = {0021-9606, 1089-7690},
  doi = {10.1063/5.0299743},
  urldate = {2026-03-17}
}

@article{zwanikken2009inflation,
  title={Inflation of the screening length induced by Bjerrum pairs},
  author={Zwanikken, Jos and Van Roij, Rene},
  journal={Journal of Physics: Condensed Matter},
  volume={21},
  number={42},
  pages={424102},
  year={2009},
  doi={10.1088/0953-8984/21/42/424102}
}

@article{adar2017bjerrum,
    author = {Adar, Ram M. and Markovich, Tomer and Andelman, David},
    title = {Bjerrum pairs in ionic solutions: A {P}oisson-{B}oltzmann approach},
    journal = {J. Chem. Phys.},
    volume = {146},
    number = {19},
    pages = {194904},
    year = {2017},
    month = {05},
    issn = {0021-9606},
    doi = {10.1063/1.4982885}
}

@misc{Dean2026,
  title = {Dielectric Response and Viscosity Due to Dipolar Interactions},
  author = {Dean, David S. and Diamant, Haim},
  year = 2026,
  publisher = {arXiv},
  doi = {10.48550/ARXIV.2604.00262},
  urldate = {2026-04-21}
}

@article{Jungwirth2018,
  title = {Ion-{{Induced Long-Range Orientational Correlations}} in {{Water}}: {{Strong}} or {{Weak}}, {{Physiologically Relevant}} or {{Unimportant}}, and {{Unique}} to {{Water}} or {{Not}}?},
  shorttitle = {Ion-{{Induced Long-Range Orientational Correlations}} in {{Water}}},
  author = {Jungwirth, Pavel and Laage, Damien},
  year = 2018,
  month = apr,
  journal = {J. Phys. Chem. Lett.},
  volume = {9},
  number = {8},
  pages = {2056--2057},
  publisher = {American Chemical Society (ACS)},
  issn = {1948-7185, 1948-7185},
  doi = {10.1021/acs.jpclett.8b01027},
  urldate = {2025-07-22}
}

@article{Laage2011,
  title = {Reorientation and {{Allied Dynamics}} in {{Water}} and {{Aqueous Solutions}}},
  author = {Laage, Damien and Stirnemann, Guillaume and Sterpone, Fabio and Rey, Rossend and Hynes, James T.},
  year = 2011,
  month = may,
  journal = {Annual Review of Physical Chemistry},
  volume = {62},
  number = {1},
  pages = {395--416},
  issn = {0066-426X, 1545-1593},
  doi = {10.1146/annurev.physchem.012809.103503},
  urldate = {2026-04-13}
}

@article{Bakker2010,
  title = {Vibrational {{Spectroscopy}} as a {{Probe}} of {{Structure}} and {{Dynamics}} in {{Liquid Water}}},
  author = {Bakker, H. J. and Skinner, J. L.},
  year = 2010,
  month = mar,
  journal = {Chemical Reviews},
  volume = {110},
  number = {3},
  pages = {1498--1517},
  issn = {0009-2665, 1520-6890},
  doi = {10.1021/cr9001879},
  urldate = {2026-04-21}
}

@article{Minh2023,
  title = {Electrical Noise in Electrolytes: A Theoretical Perspective},
  shorttitle = {Electrical Noise in Electrolytes},
  author = {Minh, Th{\^e} Hoang Ngoc and Kim, Jeongmin and Pireddu, Giovanni and Chubak, Iurii and Nair, Swetha and Rotenberg, Benjamin},
  year = 2023,
  month = feb,
  journal = {Faraday Discussions},
  volume = {246},
  pages = {198},
  doi = {10.1039/D3FD00026E},
  urldate = {2024-07-07},
}

@book{Robinson2002,
  title = {Electrolyte {{Solutions}}},
  author = {Robinson, R A and Stokes, R H},
  year = 2002,
  publisher = {Dover Publications}
}

@book{Harned1943,
  title = {The {{Physical Chemistry}} of {{Electrolyte Solutions}}},
  author = {Harned, H S and Owen, B B},
  year = 1943,
  publisher = {American Chemical Society Monograph Series}
}

@book{Resibois1968,
  title = {Electrolyte {{Theory}}: {{An Elementary Introduction}} to a {{Microscopic Approach}}},
  author = {R{\'e}sibois, P.},
  year = 1968,
  publisher = {{Harper and Row Publishers, New York, Evanston and London}}
}

@article{Yang2023,
  title = {Solvent {{Effects}} on {{Structure}} and {{Screening}} in {{Confined Electrolytes}}},
  author = {Yang, Jie and Kondrat, Svyatoslav and Lian, Cheng and Liu, Honglai and Schlaich, Alexander and Holm, Christian},
  year = 2023,
  month = sep,
  journal = {Physical Review Letters},
  volume = {131},
  pages = {118201},
  number = {11},
  publisher = {American Physical Society (APS)},
  issn = {0031-9007, 1079-7114},
  doi = {10.1103/physrevlett.131.118201},
  urldate = {2025-07-14}
}

@incollection{Sega2013,
  title = {On the Calculation of the Dielectric Properties of Liquid Ionic Systems},
  booktitle = {Recent {{Advances}} in {{Broadband Dielectric Spectroscopy}}},
  author = {Sega, Marcello and Kantorovich, Sofia S. and Arnold, Axel and Holm, Christian},
  editor = {Kalmykov, Y, P},
  year = 2013,
  pages = {103--122},
  publisher = {Springer Netherlands},
  issn = {18746500},
  doi = {10.1007/978-94-007-5012-8_8},
  isbn = {978-94-007-5011-1}
}

@article{Sega2014,
  title = {Communication: {{Kinetic}} and Pairing Contributions in the Dielectric Spectra of Electrolyte Solutions},
  author = {Sega, M. and Kantorovich, S. S. and Holm, C. and Arnold, A.},
  year = 2014,
  journal = {Journal of Chemical Physics},
  volume = {140},
  number = {21},
  pages={211101},
  doi={10.1063/1.4880237}
}

@book{Dhont1996,
  title = {An {{Introduction}} to {{Dynamics}} of {{Colloids}}},
  author = {Dhont, J K G},
  year = 1996,
  publisher = {Elsevier}
}

@book{Doi1988,
  title = {The {{Theory}} of {{Polymer Dynamics}}},
  author = {Doi, M and Edwards, S F},
  year = 1988,
  publisher = {Oxford University Press}
}

@article{Pireddu2024,
  title = {Impedance of Nanocapacitors from Molecular Simulations to Understand the Dynamics of Confined Electrolytes},
  author = {Pireddu, Giovanni and Fairchild, Connie J. and Niblett, Samuel P. and Cox, Stephen J. and Rotenberg, Benjamin},
  year = 2024,
  month = apr,
  journal = {Proceedings of the National Academy of Sciences},
  volume = {121},
  number = {18},
  pages = {e2318157121},
  issn = {0027-8424, 1091-6490},
  doi = {10.1073/pnas.2318157121},
  urldate = {2026-04-30}
}

@article{Roy2010,
  title = {Dynamics in an {{Idealized Ionic Liquid Model}}},
  author = {Roy, Durba and Patel, Nikhil and Conte, Sean and Maroncelli, Mark},
  year = 2010,
  month = jul,
  journal = {The Journal of Physical Chemistry B},
  volume = {114},
  number = {25},
  pages = {8410--8424},
  issn = {1520-6106, 1520-5207},
  doi = {10.1021/jp1004709},
  urldate = {2026-05-04}
}

@article{Schroder2008,
  title = {On the Computation and Contribution of Conductivity in Molecular Ionic Liquids},
  author = {Schr{\"o}der, C. and Haberler, M. and Steinhauser, O.},
  year = 2008,
  month = apr,
  journal = {The Journal of Chemical Physics},
  volume = {128},
  number = {13},
  pages = {134501},
  issn = {0021-9606, 1089-7690},
  doi = {10.1063/1.2868752},
  urldate = {2026-05-04}
}

@misc{dataset,
	howpublished = "\url{https://doi.org/10.5281/zenodo.23319413}"
}

\end{document}